\documentclass[a4paper,11pt]{article}
\pdfoutput=1 % if your are submitting a pdflatex (i.e. if you have
\usepackage{jheppub} % for details on the use of the package, please
\usepackage[T1]{fontenc} % if needed

\usepackage{CJK}
\usepackage{tikz}
\usepackage{amssymb}
\usepackage{epsfig}
\usepackage{bbm}
\usepackage{bbold}
\usepackage{graphicx}
\usepackage{hyperref}
\usepackage{romannum}
\usepackage{graphics,graphpap}
\usepackage{graphicx}
\usepackage{amsmath,exscale,amsthm,bm}
\usepackage{graphicx}
\usepackage{yfonts}
\usepackage{mathrsfs}
\usepackage[toc]{appendix}
\usetikzlibrary{trees}
\usetikzlibrary{positioning,arrows}
\usetikzlibrary{decorations.pathmorphing}
\usetikzlibrary{decorations.markings}
\usetikzlibrary{decorations.text}
\usetikzlibrary{fit,chains,calc,shapes.geometric,intersections}

\begin{document}
\title{{\bf Gravitational waves from first-order phase transitions in Majoron models of neutrino mass}}

\author[a]{Pasquale Di Bari,}
\author[b]{Danny Marfatia} 
\author[c,d,a]{and Ye-Ling Zhou} 

\affiliation[a]{School of Physics and Astronomy, University of Southampton \\ Southampton, SO17 1BJ, U.K.}
\affiliation[b]{Department of Physics and Astronomy, University of Hawaii at Manoa \\ Honolulu, HI 96822, USA} 
\affiliation[c]{School of Fundamental Physics and Mathematical Sciences, \\
Hangzhou Institute for Advanced Study, UCAS, Hangzhou 310024, China} 
\affiliation[d]{International Centre for Theoretical Physics Asia-Pacific, Beijing/Hangzhou, China}

\abstract{We show how the generation of right-handed neutrino masses in Majoron models may be associated with a first-order phase transition and accompanied by the production of a stochastic background of gravitational waves (GWs). We explore different energy scales with only renormalizable operators in the effective potential. If the phase transition occurs above the electroweak scale, the signal can be
tested by future interferometers. We consider two possible energy scales for phase transitions below the electroweak scale. If the phase transition occurs at a GeV, the signal can be tested at LISA and provide a complementary cosmological probe to right-handed neutrino searches at the FASER detector. If the phase transition occurs below 100 keV, we find that the peak of the GW spectrum is two or more orders of magnitude below the putative NANOGrav GW signal at low frequencies, but well within reach of the SKA and THEIA experiments. We show how searches of very low frequency GWs are motivated by solutions to the Hubble tension in which ordinary neutrinos interact with the dark sector. We also present general calculations of the 
phase transition temperature and Euclidean action that apply beyond Majoron models.}

\maketitle
\flushbottom

\def\a{\alpha}
\def\b{\beta}
\def\c{\chi}
\def\d{\delta}
\def\e{\epsilon}
\def\f{\phi}
\def\g{\gamma}
\def\h{\eta}
\def\i{\iota}
\def\j{\psi}
\def\k{\kappa}
\def\la{\lambda}
\def\m{\mu}
\def\n{\nu}
\def\o{\omega}
\def\p{\pi}
\def\q{\theta}
\def\r{\rho}
\def\s{\sigma}
\def\t{\tau}
\def\u{\upsilon}
\def\x{\xi}
\def\z{\zeta}
\def\D{\Delta}
\def\F{\Phi}
\def\G{\Gamma}
\def\J{\Psi}
\def\L{\Lambda}
\def\O{\Omega}
\def\P{\Pi}
\def\Q{\Theta}
\def\S{\Sigma}
\def\U{\Upsilon}
\def\X{\Xi}

%Varletters
\def\ve{\varepsilon}
\def\vf{\varphi}
\def\vr{\varrho}
\def\vs{\varsigma}
\def\vq{\vartheta}

\def\dg{\dagger}                                     % hermitian conjugate
\def\ddg{\ddagger}                                   % double dagger
\def\wt#1{\widetilde{#1}}                    % big tilde
\def\mt{\widetilde{m}_1}
\def\mti{\widetilde{m}_i}
\def\mtj{\widetilde{m}_j}
\def\rt{\widetilde{r}_1}
\def\mtt{\widetilde{m}_2}
\def\mttt{\widetilde{m}_3}
\def\rtt{\widetilde{r}_2}
\def\mb{\overline{m}}
\def\VEV#1{\left\langle #1\right\rangle}        % < >
\def\be{\begin{equation}}
\def\ee{\end{equation}}
\def\ds{\displaystyle}
\def\ra{\rightarrow}

\def\bea{\begin{eqnarray}}
\def\eea{\end{eqnarray}}
\def\NO{\nonumber}
\def\Bar#1{\overline{#1}}
\def\ylz{\textcolor{red}}

\section{Introduction} 

The recent discovery of GWs provides a powerful phenomenological tool to probe the imprint of new physics in the 
early universe. Primordial GWs that survive until the present could have  been produced by mechanisms that rely on new physics.  
In particular, strong first-order phase transitions  are a well known source of 
gravitational radiation~\cite{Witten:1984rs,Hogan:1986qda,Turner:1990rc} that could be
detectable in the form of  a stochastic GW background.  These have been 
studied within the standard model (SM), either in relation to a QCD phase transition~\cite{Witten:1984rs}
or to an electroweak phase transition~\cite{Kamionkowski:1993fg}. 
However, it is now established that both electroweak symmetry breaking and 
the approximate QCD chiral symmetry breaking occur 
as a smooth crossover, with no phase transition, so that no associated 
stochastic GW background is predicted within the SM. 
Still, the necessity to extend the SM
to explain outstanding cosmological puzzles and to incorporate neutrino masses and mixing,
motivates the study of first-order phase transitions in the early universe. 
In this way primordial GWs are
an important tool, complementary to more traditional ones, to probe physics beyond the SM.
For example, GW production  from  a strong first-order phase transition in the early universe has been studied  
jointly with electroweak baryogenesis   within supersymmetric models~\cite{Apreda:2001us},
or in association with dark matter genesis~\cite{Witten:1984rs,Bai:2018dxf}, or both~\cite{Huang:2017kzu,Hall:2019rld}, and even
with the simultaneous genesis of dark matter and  neutrino masses~\cite{DiBari:2020bvn}.

In this paper we explore first-order phase transitions within extensions of the
SM that can explain neutrino masses and mixing. In this respect, the type-I seesaw mechanism 
offers an attractive way to explain neutrino masses with the addition of right-handed (RH) neutrinos with large Majorana masses. 
In this case the generation of heavy neutrino Majorana masses can provide a natural framework for the realization
of a first-order phase transition in the early universe. In particular, the Majoron model~\cite{Chikashige:1980ui} provides a way to embed
the type-I seesaw mechanism and at the same time to generate Majorana masses.  This is done through
the introduction of a complex scalar field with terms in the Lagrangian respecting $U(1)_L$ symmetry
spontaneously broken below a critical temperature.   Since spontaneous symmetry breaking can occur via a strong first-order phase transition, 
an associated production of GWs is possible, in a way that  GW experiments can probe the origin of Majorana masses and give insight 
on the scale of symmetry breaking.  
The scale of symmetry breaking is not predicted by the model. It is useful to distinguish {\it high scale scenarios}, in which the scale is above the electroweak scale, from {\it low scale scenarios},
in which the scale is below the electroweak scale. In the latter
case, it is  intriguing that the pulsar timing array experiment
NANOGrav has recently found  strong evidence for a stochastic GW background in the frequency band 
$\sim 10^{-8}$~Hz in the 12.5-year dataset~\cite{Arzoumanian:2020vkk}.
In our scenario, the phase transition occurs in the dark sector which includes the complex scalar and RH neutrinos.
The dark sector interacts with the SM sector gravitationally and with nonstandard interactions. 
The possibility that a low scale dark phase transition may address the NANOGrav result has been recently 
considered in Ref.~\cite{Nakai:2020oit} (see also \cite{Bian:2021lmz} for a Bayesian analysis comparing it with other explanations) and, more specifically, within a Majoron model with nonrenormalizable 
effective operators in Ref.~\cite{Addazi:2020zcj}. A general discussion on GW production from a first-order phase transition in 
dark sectors has been presented in Refs.~\cite{Breitbach:2018ddu,Fairbairn:2019xog}.
%though this has been done only to a phenomenological level, 
%without specifying a scalar field potential whose dynamics
%would be responsible for the first-order phase phase transition 
%and eventually  for GW production. 
In our paper we consider a few scenarios with only renormalizable terms in the tree level potential
and including one-loop zero and finite temperature contributions. 
For each model we calculate the GW spectrum and compare it to the sensitivity of existing and planned 
experiments. %We also comment on uncertainties and integrated signal-to-noise calculation. 

The paper is organized as follows. In Section~2,
we discuss the form of the effective potential within different
scenarios. In particular,  we discuss both the tree-level and the one-loop, zero temperature and  finite temperature contributions to 
the effective potential in the high scale case. %These restore the symmetry at temperatures much above the critical  temperature. 
In Section~3, we discuss the first-order phase transition and the associated 
GW production for high scale scenarios.  We scan over the parameter space (at or above the electroweak scale) and show that some of the models can be tested at future gravitational interferometers. 
In Section~4 we consider low energy scale scenarios, well below the electroweak scale. 
In particular, we attempt to reproduce 
the NANOGrav signal and also make connection with the Hubble tension in the $\Lambda$CDM model .
Finally, in Section~5, we highlight some points and draw our conclusions.

%%%%%%%%%%%%%%%%%%%%%%%%%%%%
\section{Effective potential in Majoron models} 
%%%%%%%%%%%%%%%%%%%%%%%%%%%%

The Majoron model is a simple extension of the SM~\cite{Chikashige:1980ui}.
Spontaneous breaking of a global $U_L(1)$ symmetry generates a Majorana mass term for the RH neutrinos,
 which in combination with a  Dirac mass term,
generates the  light neutrino masses via the type-I seesaw mechanism. 
To implement this, the SM field content is augmented with $N \geq 2$ RH neutrinos $N_I$,  
and a complex scalar singlet,
\be
\sigma ={1 \over \sqrt{2}}\,(\s_1 + i \, \s_2) \,  ,
\ee
 where the real component is CP-even and the imaginary component  is CP-odd. 
 The new scalar $\sigma$ has a tree level potential $V_0(\sigma)$. Then, the
 tree-level extension of the SM Lagrangian is
 \bea\label{eq:L_l}
-{\cal L}_ {N_I+\s} & = & 
 \overline{L_{\a}}\,h_{\a I}\, N_{I}\, \widetilde{\Phi} 
+  {\lambda_{I}\over 2}  \, \sigma \, \overline{N_{I}^c} \, N_{I}
+ V_0(\sigma)
+ {\rm h.c.} \,,
\eea
where $\widetilde{\Phi}$ is the dual Higgs doublet.
As we will discuss, above a critical temperature $T_{\rm c}$, $\langle \s \rangle = 0$ and the RH neutrinos are massless.
Since the lepton  doublets $L_\alpha$ and RH neutrinos $N_I$ have $L = 1$, and  $\sigma$ has $L=2$,
lepton number is conserved.  Below $T_{\rm c}$, the $U_L(1)$ symmetry is broken, the scalar $\s$ gets a vacuum expectation value  
$\langle \s \rangle = v/\sqrt{2}$ and the RH neutrinos gain Majorana masses $M_I = v\, \lambda_I/\sqrt{2}$, 
which leads to lepton number violation and  small masses for the SM Majorana neutrinos via the type-I seesaw mechanism. 
We assume in this and next section that $T_{\rm c} \gtrsim T_{\rm ew} \sim 100 \, {\rm GeV}$, so that 
the Majorana mass term is not generated later than the Dirac mass term.

\subsection{Minimal model}

Consider the form of the tree-level potential $V_0(\sigma)$.  We neglect
a possible contribution from the mixing of the new scalar with the standard Higgs boson.
%We start first by considering the simplest case, the basic model, and then we discuss some variant that, as we will see, will prove important to enhance the GW signal.   
%Moreover, we can  decompose
%$V_0(\sigma)$ into a lepton conserving contribution $V_L(\sigma)$ and a lepton violating contribution
%$V_{L \!\!\!/}(\sigma)$, explicitly
%\be
%V_0(\sigma) = V_L(\sigma) + V_{L \!\!\!/}(\sigma) \,  .
%\ee
Imposing a global $U(1)_L$ symmetry and keeping only renormalizable terms, 
the tree level potential can be simply written as
\begin{eqnarray} \label{eq:V_L}
V_0(\sigma) = - \mu^2 |\sigma|^2 + \lambda |\sigma|^4 \, .
\end{eqnarray}
Here, $\lambda$ is real and positive, so that the potential is bounded from below, and  $\mu^2$ is real and positive
to ensure  the existence of degenerate nontrivial stable minima
with  $\langle \s \rangle   = v_0\,e^{i\theta}/\sqrt{2}$ ($0\leq \theta < 2\pi$), where $v_0 \equiv \sqrt{\mu^2/\lambda}$.  %Since all minima are equivalent, without loss of generality we can assume
%that the symmetry is broken along the direction $\theta =0$.
In the broken phase we can rewrite $\s$ as
\be\label{sigma}
\sigma ={e^{i\theta} \over \sqrt{2}}\,(v_0 +  S + i \, J) \,  ,
\ee 
where  $S$ is a massive  field with $m_S^2= 2 \lambda v_0^2$ 
and $J$ is  the Majoron, a massless Goldstone field. 
The vacuum expectation value of $\sigma$ also generates RH neutrino masses $M_I = \lambda_I \, v_0/\sqrt{2}$ and these, via
the type-I seesaw mechanism, lead to a light neutrino mass matrix given by the seesaw formula, %\cite{seesaw}
\begin{eqnarray} \label{eq:seesaw}
(m_\nu)_{\alpha\beta} = - {v_{\rm ew}^2 \over 2}{h_{\alpha I} h_{\beta I} \over M_I} \, ,
\end{eqnarray}
where $v_{\rm ew}$ is the standard Higgs vacuum expectation value. Equation~(\ref{eq:seesaw}) is understood to be in the flavor basis
in which the charged lepton and Majorana mass matrices are diagonal.

The tree-level  zero-temperature potential $V_0(\sigma)$ describes the broken symmetry phase of the dark sector that includes
the scalar field $\s$ and the RH neutrinos. The dark sector couples to the thermal bath of SM particles via the Yukawa interactions
of the $N_I$'s. In particular, $2 \ra 2$ scatterings (such as  $L+ q  \ra N_I + t   $) can thermalize the RH neutrinos even if they are massless. 
Indeed, the abundance of the ultrarelativistic RH neutrino $N_I$, normalized so
that $N_{N_I}^{\rm eq} = 1$, evolves simply as $N_{N_I} = 1 - \exp[-T^{\rm eq}_{I}/T]$,
where $T^{\rm eq}_I$ is the $N_I$ equilibration temperature given by \cite{Garbrecht:2013bia,DiBari:2019zcc} 
\be\label{Teq}
T^{\rm eq}_I \simeq 0.2 \, {(h^\dagger \, h)_{II} \, v_{\rm ew}^2 \over m_{\rm eq}} = 0.2 \, M_I \, K_I\,  ,
\ee  
where $m_{\rm eq} = [16\pi^{5/2}\sqrt{g^\star_\rho}/(3\sqrt{5})]\,(v_{\rm eq}/M_{\rm P}) \simeq 1.1 \,{\rm meV}\,\sqrt{g^\star_\rho/g^{\rm SM}_\rho}$ 
is the usual effective equilibrium neutrino mass, and $K_I \equiv (h^\dagger \, h)_{II} \, v_{\rm ew}^2 /(M_I \, m_{\rm eq})$ is the usual RH neutrino decay parameter. Here, $g_{\rho}^\star \equiv g_{\rho}(t_{\star})$ is the number of ultrarelativistic degrees of freedom at the time   of the phase transition $t_\star$,  which in the cases of interest to us, is just after the time of RH neutrino production.
In the standard case, when the dynamics occurs above the electroweak scale, this can be well approximated by the SM number of degrees of
freedom. More accurately, one should include the small contribution from the RH neutrinos and the complex scalar, 
but this level of accuracy is not needed at this stage. 

For $M_I/T_{\rm c} \gtrsim 5/K_I$, the RH neutrino  $N_I$ reaches thermal equilibrium before the onset of the phase transition, and is unable to 
decay prior to this because it is massless.  In order for the seesaw formula to be consistent with neutrino oscillation data without fine-tuning, at least two 
decay parameters must be in the range $10$--$100$. Only a possible third RH neutrino can have an arbitrarily small $K_I$. 
For definiteness, when studying high scale scenarios, we assume that all three RH neutrinos have decay parameters and masses that satisfy $K_I \, M_I \gtrsim 5\,T_{\rm c}$, so 
that they are all in thermal equilibrium prior to the onset of the phase transition. 
 
In addition, the $N_I$'s couple to the scalar field through $N_I+ N_I \leftrightarrow \sigma$. In this case, the interaction rate is 
 $\Gamma_{\s} \sim (M_I/v_0)^2\,T$ and the processes can be assumed to be in equilibrium (i.e., $\Gamma_{\s} > H$)
at the time of the phase transition for a reasonable choice of parameters. In the following we assume that
the dark sector is in thermal equilibrium with the SM sector prior to the phase transition.

Finite temperature effects induce symmetry restoration at temperatures above the critical temperature~\cite{Kirzhnits:1972ut} and govern the dynamics responsible for the  phase transition from the metastable vacuum (where lepton number is 
conserved and neutrinos are massless), to the true vacuum (where lepton number is violated and neutrinos acquire masses). 
These effects are  well described by the one-loop  effective potential at finite temperatures $V^T_1(\sigma)$.
The one-loop contribution also generates a zero-temperature term  $V^0_1(\sigma)$ that needs to be included for consistency.
Then, the finite-temperature effective potential at one-loop is~\cite{Dolan:1973qd}
\be
V_{\rm eff}^T(\sigma) = V_0 (\sigma) + V^0_1(\sigma) +  V^T_1(\sigma) \,  .
\ee
The  zero temperature one-loop contribution is given by the 
Coleman-Weinberg potential that can be written, using cut-off regularization, as~\cite{Dolan:1973qd,Anderson:1991zb,Dine:1992wr,Quiros:1999jp}
\bea\label{V01}
V^0_1(\sigma) & = & {1 \over 64 \, \pi^2}\,\left\{m_\s^4 (\s) \, \left(\log {m^2_\s(\s) \over m^2_\s(v_0)} - {3 \over 2}\right) + 2\,m_\s^2 (\s)\,m^2_\s(v_0) \right. \\ \nonumber
& & \left. \;\;\;\;\;\;\; - 2\,\sum_I \, \left[M_I^4(\s) \, \left(\log {M_I^2(\s) \over M^2_I(v_0)} - {3 \over 2}\right) + 2\,M^2_I (\s)\,M_I^2(v_0) \right] \right\}  \,  ,
\eea
%where $\mu_R$ is the renormalization mass scale and 
where the prefactor of two in the second line accounts for the two degrees of freedom for each of the three RH neutrinos. 
The one-loop thermal potential is given by~\cite{Anderson:1991zb,Dine:1992wr,Quiros:1999jp}
\be \label{eq:V_th}
V^T_1(\sigma)  = \frac{T^4}{2\pi^2} \left[ J_B\left(\frac{m_\s^2(\s)}{T^2}\right) -
2 \sum_I J_F\left(\frac{M_I^2(\s)}{T^2}\right) \right]  \,  ,
\ee
where the thermal functions are 
\be
J_{B,F}(x^2) = \int_0^{\infty} dy \, y^2 \, \log (1 \mp e^{-\sqrt{x^2+y^2}})\,,
\ee
 and $m_\s^2(\s)$ and $M_I^2(\s)$ are the shifted masses.
Without loss of generality, we set $\theta = 0$ in Eq.~(\ref{sigma}) so that the minimum of the potential including thermal effects, 
lies along the real axis of $\s$. In this way we can track the dynamics of the potential as a function of $\s_1$.
The tree-level potential becomes
\be
V_0(\s_1) = -{1\over 2}\,\mu^2 \, \s_1^2 + {\lambda\over 4}\,\sigma_1^4 \,  ,
\ee
and the shifted masses are
\be
m_\s^2(\s_1) \equiv {d^2V^0(\sigma_1)\over d^2 \sigma_1}  = - \lambda v_0^2  + 3 \lambda \s_1^2 
\ee
and
\be
M_I^2(\s_1) = \lambda_I^2 \, {\s_1^2 \over 2} \,  .
\ee
Note that $m_\s^2(\sigma_1)$ can be either positive or negative, depending on the size of $\sigma_1$. 
An imaginary part of the effective potential is obtained for $m_\s^2(\s_1)<0$. This corresponds to decay widths of 
modes expanded around unstable regions of field space and does not affect the computation of the phase transition~\cite{Delaunay:2007wb}. Therefore, in the following  we neglect the imaginary part 
of the potential. We account for resummed thermal masses 
 by replacing the tree-level shifted mass in Eq.~(\ref{eq:V_th})  by~\cite{Parwani:1991gq}
\be\label{resummation}
m_\s^2(\s_1) \to {\rm m}_{\s , T} ^{2}(\s_1) = m_\s^2(\s_1)    + \Pi_\s\,.
\ee
 The leading contribution to the thermal mass, i.e., the Debye mass, is given by
\be
\Pi_\s = \left( \frac{2+d_{\rm scalar}}{12} \lambda + N\,\frac{M^2}{24 v_0^2} \right) T^2 \,  ,
\ee 
 where $d_{\rm scalar} = 2$ for a complex scalar. This yields the 
 so-called {\it dressed effective potential}~\cite{Curtin:2016urg,Croon:2020cgk}.
 $M$ denotes either the mass of the heaviest RH neutrino, if it gives the dominant contribution (in which case $N= 1$), or 
a common mass for quasi-degenerate RH neutrinos (in which case $N$ is  the number of RH neutrinos).
This allows us to reduce the number of parameters while spanning the space between $N=1$ (hierarchical RH neutrinos) and $N=3$ (quasi-degenerate RH neutrinos).

 It is useful to write the effective potential in terms of the high temperature expansion of the thermal functions~\cite{Anderson:1991zb,Quiros:1999jp},
 \begin{eqnarray}
J_B(x^2) & = & -\frac{\pi ^4}{45} +\frac{\pi^2 x^2}{12} -\frac{\pi}{6}  (x^{2})^{3/2}-\frac{x^4}{32}\,\log {x^2 \over a_B} + {\cal O}(x^6) \,, \nonumber\\
J_F(x^2) & = & \frac{7 \pi ^4}{360} -\frac{\pi^2 x^2}{24} - \frac{x^4}{32} \, \log {x^2 \over a_F} + {\cal O}(x^6)  \, ,
\end{eqnarray}
where $a_B = 16\,\pi^2\,\exp(3/2 -2\gamma_E)$ and $a_F = \pi^2 \exp(3/2 -2\gamma_E)$, with $\gamma_E = 0.5772$. 
We checked that this expansion is accurate to better than $\sim 5\%,12\%,27\%$ for $x^2 \leq 3,4,5$, respectively.  In our numerical work, we set $x^2 \leq 3$.

 Since we are assuming that the RH neutrinos
 are thermalized, the field dependence in the logarithmic term  in Eq.~(\ref{V01})
 cancels the logarithmic term in the high temperature expansion of $V^T_1(\sigma) $ to leave the
one-loop finite-temperature effective potential in a polynomial form,
 \be\label{VTeffminimal}
 V^T_{\rm eff}(\s_1) \simeq D\, (T^2 - T_0^2) \s_1^2 - A \, T \, \s_1^3 + \frac{1}{4}\lambda_T\, \s_1^4 \,  ,
 \ee
where the  destabilization temperature $T_0$ is defined by %%
\be
2\,D\,T_0^2 =   \lambda\,v_0^2 +{N \over 8\,\pi^2}\,{M^4 \over v_0^2} 
-{3\over 8 \,\pi^2}\lambda^2 \, v_0^2  \,  ,
\ee
and the dimensionless coefficients $D$ and $A$ are
\be\label{DA}
D = {{\lambda \over 8} + {N\over 24}\,{M^2 \over v_0^2}} \,   \;\;\;\; \mbox{\rm and} \;\;\;\;
A =  {(3\,\lambda)^{3/2} \over 12\pi } \,  \,  .
\ee
 The cubic term  is obtained using the approximation, 
$m^2_{\s}(\s_1) = \Pi_\s -\lambda v_0^2+3\lambda \s_1^2 \simeq 3\lambda \s_1^2$. This approximation
works quite well since the cubic term gives a nonnegligible contribution 
only when the field is large. The dimensionless temperature dependent 
coefficient $\lambda_T$ is given by
\be\label{lambdaT}
\lambda_T = \lambda  - \frac{N\, M^4}{8\,\pi^2 \, v_0^4}  \, \log {a_F \, T^2 \over e^{3/2}\,M^2} 
+  {9\lambda^2 \over 16 \pi^2}\,\log {a_B \, T^2 \over e^{3/2}\,m_S^2} \,  .
\ee
In the following it will prove convenient to define 
\be
\widetilde{M}_T^2\equiv 2\,D\,(T^2 - T_0^2) \,   ,
\ee
so that  the thermal effective potential can be written more compactly as
\be\label{VTeffminimal2}
 V^T_{\rm eff}(\s_1) \simeq {1\over 2}\, \widetilde{M}_T^2\,\s_1^2 - A \, T \, \s_1^3 + \frac{1}{4}\lambda_T\, \s_1^4 \,  .
 \ee
Finally, note that the massive scalar $S$ must decay for its thermal abundance not to overclose the universe.
Since the scalar and RH neutrino masses are of roughly the same order-of-magnitude as $v_0$, we can always assume that 
the mass of the lightest RH neutrino $M_1 < m_S$ and that $S$ decays quickly enough via $S \ra N_1 + N_1$. 

%%%%%%%%%%%%%%%%%%%%%%%%%%%%%%%%%%%%%%%%
\subsection{Explicit symmetry breaking term and a massive Majoron}
%%%%%%%%%%%%%%%%%%%%%%%%%%%%%%%%%%%%%%%%

It might be useful for various applications to give the Majoron a mass.  
This can be done by adding an explicit symmetry breaking term $ V_{L \!\!\!/}(\sigma)$ to the tree-level potential, 
obtaining
\be\label{treelevel}
V_0(\sigma) =  - \mu^2 |\sigma|^2 + \lambda |\sigma|^4+ V_{L \!\!\!/}(\sigma) \,  .
\ee
Consider the $U(1)_L$-breaking renormalizable term,
\begin{eqnarray} \label{eq:V_LB}
V_{L \!\!\!/}(\sigma) = - \sqrt{2} \, \widetilde{m} \, \sigma^2 \sigma^* + {\rm h.c.}\, ,
\end{eqnarray}
where the mass-dimension parameter $\widetilde{m}$ can always be taken to be real by absorbing a phase into $\sigma$. 
Other renormalizable terms that break $U(1)_L$ explicitly are proportional to $\sigma^2$, $\sigma^3$, $\sigma^4$ and $\sigma^3\sigma^*$ and their complex conjugates. However, these terms are invariant under a discrete symmetry, either $Z_2$, $Z_3$ or $Z_4$, and
the spontaneous breaking of a discrete symmetry may induce a domain wall problem~\cite{Vilenkin:1984ib,Kibble:1976sj}.\footnote{Domain walls form after spontaneous symmetry breaking. The tension of these walls is estimated to be $\sim m_J v^2$, which should be lower than MeV$^3$  to be consistent with the primordial density fluctuations,  $\delta \rho / \rho \lesssim {\cal O}(10^{-5})$ \cite{Zeldovich:1974uw}. Consequently, the explicit $U(1)$ breaking parameter is constrained to be tiny: $|\tilde{m}|/v \sim m_J^2 /v^2 \lesssim {\rm MeV}^6/v^6$.} 
For this reason we assume $V_{L \!\!\!/}$ in Eq.~(\ref{eq:V_LB}) 
to be the only explicit breaking term.  Its inclusion  breaks the degeneracy of the minima
with only one minimum located on the real axis at $\theta = 0$ for $\widetilde{m} > 0$,\footnote{Alternatively, for $\widetilde{m} < 0$ the minimum is located at $\theta = \pi$; the two options are fully equivalent.}
%Since they are fully equivalent, without loss of generality, we can consider the case $\widetilde{m}, v_0 > 0$. 
and with a vacuum expectation value,
\begin{eqnarray}
v_0(\widetilde{m})= \sqrt{\frac{\mu^2}{\lambda} + \left( \frac{3\,\widetilde{m}}{2\lambda} \right)^2} + \frac{3\,\widetilde{m}}{2\,\lambda}  \,  .
\end{eqnarray}
Thus, the tree-level mass of $S$ gets modified and $J$ becomes a pseudo-Goldstone boson by acquiring a mass due to the explicit breaking term:
\be
m_S^2 = 2\lambda v_0^2 - 3 \widetilde{m} v_0 \,  , \;\;\;
m_J^2 = \widetilde{m}\, v_0\, .
\ee

We now consider how the potential gets modified by the introduction of the explicit symmetry breaking term.
As for the minimal model, the minimum lies on the real axis and potential can be written in terms of $\sigma_1$.
 The tree level potential is
\be
V_0(\s_1) = -{1\over 2}\,\mu^2 \, \s_1^2  - \widetilde{m}\,\s_1^3  + {\lambda\over 4}\,\sigma_1^4\,  ,
\ee
which yields the shifted mass,
\be
m_\s^2(\s_1, \widetilde{m}) \equiv {d^2V^0(\sigma_1)\over d^2 \sigma_1}  = - \lambda v_0^2 + 3 \widetilde{m} v_0  - 6 \widetilde{m}\s_1 + 3 \lambda \s_1^2  \,  .
\ee
Again, we neglect the imaginary part  of the potential and
account for the resummed thermal masses in the effective potential by the replacement,
\be
m_\s^2(\s_1,\widetilde{m}) \to {\rm m}_{\s , T} ^{2}(\s_1,\widetilde{m}) 
= m_\s^2(\s_1,\widetilde{m})  + \Pi_\s   \,  .
\ee
Assuming that the RH neutrinos get thermalized, the explicit symmetry breaking term perserves the polynomial form of the
one-loop finite temperature effective potential,
%the field dependence in the logarithmic term  in the zero-temperature one-loop contribution (\ref{V01})
% cancels out with the logarithmic term in the high temperature expansion of $\Delta V^T_1(\sigma) $ and one obtains that with the addition of the explicit symmetry breaking term the one-loop finite temperature effective potential is still in a polynomial form:
 \be
 V^T_{\rm eff}(\s_1,\widetilde{m}) = - {\widetilde{m}\over 4}\,\left({3\,m^2_S\over 4\,\pi^2} + f(T) +T^2 \,\right)\,\sigma_1 + D\, (T^2 - T_0^2) \s_1^2 - 
 \, [A \, T+\widetilde{m}(1+g(T))]  \, \s_1^3 + \frac{1}{4}\lambda_T\, \s_1^4 \,  .
 \label{VTexp}
 \ee
The destabilization temperature now receives an additional contribution and becomes
\be
T_0^2(\widetilde{m}) =  T_0^2(\widetilde{m} =0) 
 -{3\,\widetilde{m}\,v_0 \over 2 \, D} = T_0^2(\widetilde{m} =0) - {3\,m^2_J \over 4 \,D } \,  .
\ee
The dimensionless coefficients $D$, $A$ and $\lambda_T$ are unchanged from the $\widetilde{m}=0$ case, while 
 the new dimensionless temperature dependent coefficients $f(T)$ and $g(T)$ are given by
\be
f(T)=  {3 \over 32\,\pi^2}\,(3m^2_J - m^2_S)\, \log {a_B \, T^2 \over e^{3/2}\,m_S^2}   \,  , 
\ee
and
\be
g(T) = {9\,\lambda \over 16 \pi^2} \, \log {a_B \, T^2 \over e^{3/2}\,m_S^2} \,  .
\ee
Note that the explicit symmetry breaking term results in a linear term and a cubic term at zero temperature in Eq.~(\ref{VTexp}).  The latter is known to be able to strengthen the phase transition but the
former is potentially dangerous and can jeopardize the phase transition if it dominates over the other terms at 
high temperatures. It is $\propto T^2$ and at high temperatures shifts the minimum
from $\sigma_1 = 0$ to $\sigma_1 \simeq \widetilde{m}/(8\,D)$. Thus,  the symmetry is not restored at high temperatures. 
This is not necessarily a problem since the universe could start from this minimum and then 
tunnel to the true minimum. The real problem is that for large enough values of $\widetilde{m}$,
the linear term dominates over the cubic term thereby removing the barrier so that there is no first-order phase
transition and, therefore, no GW production.  In fact, an upper bound on  $\widetilde{m}$ is obtainable by requiring that a first-order phase transition occurs.  We have not
determined this upper bound but do find that there is no first-order phase transition for $\widetilde{m} \gtrsim 10^{-4}\,v_0$. We do not pursue this scenario and set $\widetilde{m} = 0$.

\subsection{Adding an auxiliary real scalar}

It is well known that the addition of a real scalar to the SM extends the parameter space in which 
a strong first-order phase transition can occur~\cite{Choi:1993cv}.
Specifically, great attention has been devoted to the possibility of successful baryogenesis 
 for values of the Higgs mass above the upper bound $m_H \lesssim 70 \, {\rm GeV}$ needed to have a first-order phase transition in the 
 SM. It has also been noticed that 
the addition of a real scalar  can enhance the GW signal produced from the phase transition~\cite{Kehayias:2009tn}.
Motivated by the positive aspects of the  simultaneous presence of two scalars, we  augment the Majoron model
with a real scalar $\eta$. New renormalizable terms contribute to the tree-level potential,
\bea
V_0(\s, \eta) = V_0(\s) + V_{\eta\s}(\s,\eta) + V_{\eta}(\eta) \,  ,
\eea
where $V_0(\s)$ is given by Eq.~(\ref{treelevel}), 
\be
V_{\eta}(\eta) = {\gamma_2 \over 2}  \, \eta^2  + {\gamma_3 \over 3} \,  \eta^3 +{\gamma_4 \over 4} \, \eta^4 \, ,
\ee
and the mixing term is
\be
V_{\eta\s}(\s,\eta) = {\delta_1 \over 2}  \, |\sigma|^2 \, \eta + {\delta_2 \over 2} \,  |\sigma|^2 \, \eta^2 \,.
\ee

Without loss of generality, we can again consider the $U(1)_L$ symmetry to be broken along the real part of $\s$. 
In the decoupling limit, with a very heavy real scalar and a small mixing angle between the two mass eigenstates at zero temperature, and replacing $\eta$ with its vev\footnote{This means that $\eta$ itself undergoes a phase transition and settles to its true vacuum prior to the
$\s_1$ phase transition.}, the thermal effective potential for $\sigma_1$ takes the form~\cite{Kehayias:2009tn},
\be\label{VTeffmut}
 V^T_{\rm eff}(\s_1,\widetilde{\mu}) \simeq {1\over 2}\, \widetilde{M}_T^2\, \s_1^2 - (A \, T  + \widetilde{\mu})\, \s_1^3 + 
 \frac{1}{4}\lambda_T\, \s_1^4 \,  .
\ee
Compared to Eq.~(\ref{VTeffminimal}) for the minimal model, there is now a cubic term
at zero temperature proportional to the parameter $\widetilde{\mu} = \d^{\, 2}_2\,v_0/(2 \, \gamma_4)$. 
The other parameters also get modified by the inclusion of the real scalar, so that there are changes to the expressions
for $A,D,\lambda_T$ and $T_0$ given for the minimal model. However,
there is always a choice of the parameters $\d_i$ and $\gamma_i$ such that the corrections are small. In sum, the primary effect of adding a
real scalar is the appearance of a zero-temperature cubic term which, as we show in the next section,
can greatly enhance the GW signal, as in the case of electroweak symmetry breaking.

\subsection{Nonthermal RH neutrinos}

So far we have assumed that all three RH neutrinos are in thermal equilibrium at the time of the phase 
transition, which is a reasonable assumption for two of the neutrinos participating in the seesaw mechanism.
However, it is possible that a third RH neutrino has very small Yukawa couplings and does not get thermalized.
For definiteness, suppose that this is the lightest, $N_1$, and that its nonthermal 
abundance $N_{N_1}$ is in general nonvanishing. Then its contribution to 
the one-loop thermal potential will be $N_{N_1}$ times the contribution of the other two RH neutrinos.
This results in a finite temperature effective potential given by
\be
 V^T_{\rm eff}(\s_1,N_{N_1}) =  {1\over 2}\, \widetilde{M}_T^2\, \s_1^2 - \, A \, T  \, \s_1^3 + \frac{1}{4}\lambda_T\, \s_1^4 + (N_{N_1} - 1) \, {M^4 \, \s_1^4 \over 32 \pi^2 \, v_0^4 } \, \log {M^2 \, \s_1^2\over a_F \, T^2 \, v_0^2} \,   .
 \ee

\section{First-order phase transition and GW production}
%%%%%%%%%%%%%%%%%%%%%%%%%%%%%%%%%%

We now consider the behavior of the effective thermal potentials as the falling temperature first approaches $T_{\rm c}$ and then $T_0$. 
We begin with a general discussion that holds for all models and then consider the four specific
potentials of the previous section.

\subsection{General description of a first-order phase transition}

%There is a particular temperature $T_1$ corresponding to a time when a second stable minimum forms at  much higher values of $\s_1$.
When a second stable minimum forms at a nonzero value of $\s_1$, the temperature is $T_1$.
As the temperature drops below $T_1$ the two minima coexist and are 
separated by a barrier, the crucial feature characterizing a first-order phase transition.
The time when the two minima are degenerate
 defines the critical temperature $T_{\rm c}$. Below the critical temperature the second minimum at nonzero $\s_1$
becomes the true minimum and tunneling between the two minima can occur in the form of nucleation of
expanding true vacuum bubbles. The nucleation probability per unit time and per unit volume in terms of the Euclidean action $S_E$ is 
$\Gamma(\s_1,T) = \Gamma_0(T)\,e^{-S_E(\s_1,T)}~$\cite{Coleman:1977py}. 
At finite temperatures $S_E(\s_1,T) \simeq S_3(\s_1,T)/T$ and $\Gamma_0(T) \simeq T^4 \, [S_3(T)/(2\pi\,T)]^{3/2}$~\cite{Linde:1981zj}, 
where $S_3$ is the spatial Euclidean action given by
\be\label{S3T}
S_3(\s_1,T) = \int\,d^3x \, \left[ {1 \over 2}\,(\bm{\nabla} \s_1)^2 + V^T_{\rm eff}(\s_1) \right] 
= 4\pi \, \int_0^\infty \,dr \, r^2 \left[ {1 \over 2}\,\left({d\sigma_1^2 \over dr^2}\right)^2 + V^T_{\rm eff}(\s_1) \right] \,  .
\ee
The physical solution for $\s_1$ minimizes $S_3(\s_1,T)$ and therefore
satisfies the equation of motion,
\be\label{EoM}
{d^2\s_1 \over dr^2} + {2 \over r}\, {d\s_1 \over dr} = {d V^T_{\rm eff}(\s_1)   \over dr} \,,
\ee
with boundary conditions $(d\s_1/dr)_{r=0} = 0$ and $\s_1(r \ra \infty) = 0$.
For $T \geq T_{\rm c}$, the nucleation probability vanishes so $S_E \ra \infty$,
and for $T \ra T_0$, $S_E \ra 0$, so that at $T_0$ all space will have nucleated and the phase transition ends. 
Between $T_c$ and $T_0$, which mark the beginning and the end of the phase transition, respectively,
a nucleation temperature $T_n$ can be defined as the temperature at which one bubble is nucleated in one Hubble volume.
The corresponding time is $t_n$, i.e., $T_{\rm n} \equiv T(t_{\rm n})$, and $\int_0^{t_{\rm n}} dt (\Gamma/H^3) = 1$.
It is also customary to define the percolation temperature as the temperature at which the fraction of space still in the false vacuum is $1/e$.{\footnote{Alternatively, the percolation temperature is defined as the temperature at which the fraction of space converted to the true vacuum is  $1/e$. }} We identify this temperature with the temperature of the phase transition $T_\star$ and the corresponding time as $t_\star$ with $T_{\star} \equiv T(t_{\star})$. 

%As we discuss in the next subsection, the value of $T_{\star}$ and of the derivative of the Euclidean action at $T_\star$ are crucial for the calculation of the GW spectrum. 
The fraction of space filled by the false vacuum at time $t$  is given by~\cite{Guth:1979bh,Guth:1981uk} 
\be
P(t) = e^{- I(t)} \,  ,
\ee
where 
\be\label{It}
I(t) = {4\pi \over 3} \, \int_{t_{\rm c}}^t \, dt' \,  \Gamma(t') \, a^3(t') \, \left[\int_{t'}^t \, dt'' \, {v_{\rm w} \over a(t'')} \right]^3 \,  ,
\ee
$a(t)$ is the scale factor and $v_{\rm w}$ is the bubble wall velocity. Then, since $P(t_\star) = 1/e$ corresponds to $I(t_\star) = 1$, the following equation for $S_E(T_\star)$ 
can be derived~\cite{Megevand:2016lpr}:
\be\label{SETstar}
S_E(T_\star) - {3\over 2} \, \log {S_E(T_\star) \over 2\pi}   =
  4 \log{T_\star \over H_\star}  - 4\,\log[T_\star \, S'_E(T_\star)]  + 
  \log(8\,\pi \, v^3_{\rm w}) \,  ,
\ee
where $H_{\star} = H(t_{\star})$.
Thus, a solution for $S_E(T_\star) \simeq S_3(T_\star)/T_\star$ is found if 
 the derivative $S'_E(T_\star)$ is known. The latter is related to the quantity $\beta/H_{\star}$
defined in the next subsection. 
If the term containing the derivative is negligible, then using $v_{\rm w} \simeq 1/\sqrt{3}$ and estimating
the logarithm of the Euclidean action at $T_{\star}$ by iteration, 
we find the numerical approximation,
\be\label{S3TH}
{S_3(T_\star)\over T_\star}  \simeq{S_3(T_H)\over T_H} \simeq 153 - 4\, \log\left[\sqrt{g_\rho^\star \over 106.75} \, {T_H \over 100\,{\rm GeV}} \right] \,  ,
\ee
where $T_H$ is defined by $\G(T_H) = H^4(T_H)$. Strictly speaking, 
$T_H > T_{\rm n} > T_\star$, but  $T_H \simeq T_\star$ to within 20\%. 
Equating this  with  Eq.~(\ref{S3T}) for a specific effective potential yields an expression for $T_{\star}$. 
Note that $T_H$ is defined so that $S_E(T_H)$ depends only on $g_{\rho}^\star$ and is independent of the effective potential. In our calculations, we solve Eq.~(\ref{SETstar}) for the greater accuracy it affords.  In Appendix~\ref{app1} we give some details on how to derive Eq.~(\ref{SETstar}) and calculate $T_\star$ using different procedures depending on the required accuracy.

\subsection{Calculation of the GW spectrum}

If we define the rate of variation of $\Gamma$ as $\beta \equiv \dot{\Gamma}/\Gamma$, then $\beta^{-1}$ gives the time scale of the phase transition. If this is much shorter than the Hubble time, then 
$\beta \simeq -(dS_E/dt)_{t_{\star}}$ and
\be\label{betaoverH}
{\beta \over H_{\star}} \simeq T_{\star} \left.{d(S_3/T) \over dT}\right|_{T_{\star}} \,  .
\ee
 This is one of two parameters that characterize the GW spectrum during the phase 
transition. It requires both the temperature of the phase 
transition and  the temperature derivative of the Euclidean action at the phase transition.

The second parameter is the strength of the transition $\a$, given by the ratio of the latent heat $\ve$ released during the phase transition 
and the energy density of the plasma at the time of the phase transition:
\be\label{alpha}
\a \equiv {\ve(T_{\star}) \over \rho(T_{\star})} \,  .
\ee
As usual, the energy density is calculated in terms of $g_{\rho}(T)$, 
and is given by $\rho(T) = g_{\rho}(T) (\pi^2 /30)\,T^4$. It receives contributions from the SM sector and from the dark sector: $g_{\rho}(T) = g_{\rho}^{\rm SM}(T) + g_{\rho}^{\rm dark}(T)$.  In the high scale scenario, $g_{\rho}^{\rm SM}(T_\star) =106.75$, and 
\be
g_{\rho}^{\rm dark}(T_\star) = g_\rho^{\rm scalars} + {7 \over 4} \, N \,  ,
\ee
where $g_\rho^{\rm scalars}$ is either 2 or 3 depending on whether or not the auxiliary real scalar $\eta$ is included in the model. The latent heat is 
\be
\ve(T_{\star}) = -\Delta V^{T_\star}_{\rm eff}(\s_1) - T_{\star} \, \Delta s(T_{\star}) =  
-\Delta V^{T_\star}_{\rm eff}(\s_1)  + T_{\star} 
\left.{\partial \Delta V^{T_\star}_{\rm eff}(\s_1) \over \partial T}\right|_{T_{\star}} \,  ,
\ee
where $\Delta V^{T_\star}_{\rm eff}(\s_1) = V^{T_\star}_{\rm eff}(\s^{\rm true}_1) - V^{T_\star}_{\rm eff}(\s^{\rm false}_1)$, and in the first relation, from thermodynamics, $\D s$ is the entropy density variation and the free energy of the system has been identified with the effective potential. Recall that in our case, $V^{T_\star}_{\rm eff}(\s^{\rm false}_1)=0$.

The GW spectrum is defined as
\be
h^2\,\O_{{\rm GW}0}(f)=  {1 \over \rho_{{\rm c}0}h^{-2}} \,  {d\rho_{{\rm GW}0}\over d\ln f} \,  ,
\ee 
where $\rho_{{\rm c}0}$ and $\rho_{{\rm GW}0}$ are respectively, the critical energy density and 
the  energy density of GWs produced during the phase transition and evolved to the present time. 
We assume that the phase transition occurs in the detonation regime with supersonic bubble wall 
velocities, i.e., $v_{\rm w} \geq c_{\rm s} = {1/\sqrt{3}}$.
This regime is approximately realized for $\a \lesssim 0.1$. Also, the duration of the phase
transition $\b^{-1}$ is quite short so that the approximation in Eq.~(\ref{betaoverH}) works quite well,
and typically $\beta/H_{\star} \gtrsim 100$ \cite{Ellis:2020awk}.  Moreover, in this regime GWs are primarily sourced
by sound waves in the plasma, so that $h^2 \, \O_{{\rm GW} 0}(f) \simeq h^2 \, \O_{\rm sw}(f)$. 
An analytic fit to numerical simulations, valid for $\a \lesssim 0.1$, yields~\cite{Caprini:2015zlo,Caprini:2019egz}
\be\label{omegasw}
h^2\Omega_{\rm sw}(f) =2.59\times 10^{-6} \, \frac{v_{\rm w}(\a)}{\beta/H_\star} \left[\frac{\kappa(\a)\, \alpha}{1+\alpha}\right]^2  
\,\left( \frac{106.75}{ g_\rho^\star} \right)^{1/3} S_{\rm sw} (f) \,,
\ee
where the spectral shape function is
\begin{eqnarray}
S_{\rm sw} (f) = \left(\frac{f}{f_{\rm sw}}\right)^3 \left[\frac{7}{4+3({f/f_{\rm sw}})^2} \right]^{7/2} \,  ,
\end{eqnarray} 
and the peak frequency is given by
\begin{eqnarray} \label{fpeak}
f_{\rm sw} =1.92\times 10^{-2}\,{\rm mHz} \, \frac{1}{v_{\rm w}} \frac{\beta}{H_\star}  \frac{T_\star}{\rm 100\,GeV} \left( \frac{g_\rho^\star}{106.75} \right)^{1/6} \, .
\end{eqnarray}
Here we normalized the number of degrees of freedom to the SM value since we are discussing phase transitions at or above the electroweak scale.
The efficiency factor $\kappa(\a)$ measures how much of the vacuum energy is converted to bulk kinetic energy. 
%This factor is correlated with $\alpha$ and the bubble wall velocity $v_{\rm w}$.
We adopt Jouguet detonation solutions since we assume that the plasma velocity behind the bubble wall
is equal to the speed of sound. Then, the efficiency factor is~\cite{Steinhardt:1981ct} 
\begin{eqnarray}
\kappa(\a) \simeq \frac{\sqrt{\alpha}}{0.135+\sqrt{0.98+\alpha}} \,,
\label{efficiencyfactor}
\end{eqnarray}
and the bubble wall velocity is $v_{\rm w}(\a) = v_{\rm J}(\a)$, where
\begin{eqnarray} \label{eq:Jouguet}
v_{\rm J}(\a) \equiv \frac{\sqrt{1/3} + \sqrt{\alpha^2 +2\alpha/3}}{1+\alpha}\,.
\end{eqnarray}
Note that Jouguet solutions provide a simple and useful prescription, but within a rigorous treatment, the bubble velocity deviates from $v_{\rm J}(\a)$, and the efficiency factor is a function of both $\a$ and $v_{\rm w}$. Moreover, the friction exerted by the plasma on the bubble wall also needs to be taken into account, leading to a much more complicated description that requires numerical solutions of the Boltzmann equations~\cite{Espinosa:2010hh}. 

There are several other theoretical uncertainties that 
may lead to significant corrections. 
Equation~(\ref{omegasw}) assumes that the duration for the bulk motion of the fluid $\gtrsim H_{\star}^{-1}$. However, if the
lifetime of the sound waves $\tau_{\rm sw}$ is less than a Hubble time, then the peak amplitude of the GW power spectrum may be suppressed by a factor, $1- 1/\sqrt{1+ 2H_{\star} \tau_{\rm sw}}$, which decreases with the strength of the phase transition~\cite{Caprini:2019egz,Ellis:2018mja,Guo:2020grp}. 
Suppression factors of ${\cal{O}}(0.1)$ have been obtained~\cite{Ellis:2018mja,Guo:2020grp},
but a precise calculation is  model dependent. % If we apply their results
%for the singlet scalar extension of the SM to our models with an auxiliary scalar, for our benchmark points the suppression factor should be
%in the range $\U = 2$--$5$.
The calculation of the
effective thermal potential relies on a few approximations and prescriptions~\cite{Croon:2020cgk}. 
In particular, it relies on a high temperature expansion of the thermal functions, a perturbative 
calculation\footnote{See Ref.~\cite{Gould:2021oba} for a discussion on the necessity of a two-loop calculation to
recover renormalization scale independence at high temperatures.}
and the prescription Eq.~(\ref{resummation}) for the resummation of thermal masses. A more accurate calculation 
might then correct our results by typically suppressing the signal. Nevertheless, some corrections are expected to enhance the GW signal.
It has  recently been found that density fluctuations with scales $H_\star^{-1 } \gtrsim \lambda_{\star} \gtrsim \beta^{-1}$ 
can significantly enhance the GW signal 
for $\beta/H_\star \gtrsim 100$~\cite{Jinno:2021ury}, as occurs in the case of phase transitions in the dark sector. In Ref.~\cite{Nakai:2020oit}, it has been pointed out that since the fraction of latent heat converted to GWs is not related to $\alpha$, but to the  dynamics in the dark sector, the 
efficiency factor in Eq.~(\ref{efficiencyfactor}) need not apply and can even be of order unity.
 From a pessimistic standpoint, our results for the peak amplitude (using  Eq.~\ref{omegasw}) may be regarded as upper bounds.

\subsection{Calculation of the Euclidean action in Majoron models and beyond}

We now specialize to the three models of the previous section. For each of them
we find an expression for the Euclidean action as a function of temperature.

\subsubsection{Minimal model}

The minimal model is characterized by the effective potential $V^T_{\rm eff}(\s_1)$ in Eq.~(\ref{VTeffminimal}).
Following the procedure of Ref.~\cite{Dine:1992wr}, we introduce the dimensionless field $\overline{\s}_1$ defined  by
\be\label{barsigma1}
\s_1 = \overline{\s}_1 \, {\widetilde{M}_T^2 \over 2\,A\,T}   \,   ,
\ee
%where $\widetilde{M}^2(T) \equiv 2\,D\,(T^2 - T_0^2)$,  
and obtain the dimensionless effective potential,
\be
\overline{V}^T_{\rm eff}(\overline{\s}_1,a) \equiv 4\,{A^2\,T^2 \over \widetilde{M}_T^6}\, V^T_{\rm eff}(\overline{\s}_1) = 
{1 \over 2}\,\overline{\s}_1^2 - {1 \over 2}\,\overline{\s}_1^3 + {a \over 8}\,\overline{\s}_1^4 \,  ,
\ee 
which depends on a single parameter,
\be
a \equiv {\lambda_T \, \widetilde{M}_T^2 \over 2\,A^2 \, T^2} \,  .
\ee
In terms of $\bar{\s}_1$, the equation of motion in Eq.~(\ref{EoM})  becomes
\be\label{EoM2}
{d^2\overline{\s}_1 \over d\overline{r}^2} + {2 \over \overline{r}}\, {d\overline{\s}_1 \over d\overline{r}} = {d \overline{V}^T_{\rm eff}(\overline{\s}_1)   \over d\overline{r}} \,   , 
\ee
where $\overline{r} = r\,\widetilde{M}_T$ is a dimensionless radial coordinate. Solving the differential equation and integrating over $d\overline{r} = \widetilde{M}_T \, d r$,
we find the Euclidean action to be
\be\label{euclidean}
{S_3 \over T} = {\widetilde{M}_T^3 \over A^2 \, T^3} \, f(a) \,  ,
\ee
where 
\be
f(a)  \simeq 4.85\,\left[1 +{a \over 4}\,\left(1 +{2.4 \over 1-a} + {0.26 \over (1-a)^2} \right)\right] \,   
\ee
provides an accurate analytical fit. 

\subsubsection{Quartic potential with zero-temperature cubic term}

This result can be easily extended to the case of a
quartic effective potential with a zero-temperature cubic term (see Eq.~\ref{VTeffmut}) obtained by adding
a real scalar to the minimal model. We define a dimensionless effective thermal potential,
\be
\overline{V}^T_{\rm eff}(\overline{\s}_1,\widetilde{a}) \equiv 4\,{\widetilde{\mu}_T^2 \over \widetilde{M}_T^6}\, V^T_{\rm eff}(\overline{\s}_1,\widetilde{\mu}) = 
{1 \over 2}\,\overline{\s}_1^2 - {1 \over 2}\,\overline{\s}_1^3 + {\widetilde{a} \over 8}\,\overline{\s}_1^4 \,  ,
\ee 
where   
\be
\widetilde{\mu}_T \equiv A\,T + \widetilde{\mu} \,  , \;\;\;\;   \widetilde{a} \equiv {\lambda_T \, \widetilde{M}_T^2 \over 2\, \widetilde{\mu}_T^2} \,  ,
\ee
and the dimensionless field $\overline{\s}_1$ is
\be\label{barsigma1b}
\s_1 = \overline{\s}_1 \, {\widetilde{M}_T^2 \over 2\,\widetilde{\mu}_T}   \,   .
\ee
The Euclidean action of Eq.~(\ref{euclidean}) gets generalized to
\be\label{euclidean2}
{S_3 \over T} = {\widetilde{M}_T^3 \over   \widetilde{\mu}_T^2\, T} \, f(\widetilde{a}) \,  .
\ee

\subsubsection{Quartic potential with both zero-temperature cubic and logarithmic field-dependent terms}

We extend this treatment to the most general effective potential which includes the zero-temperature cubic term (generated by the addition of a real scalar), and a logarithmic field-dependent term (that arises if one RH neutrino has a nonthermal abundance):
\be
 V^T_{\rm eff}(\s_1,\widetilde{\mu},\lambda_1) =  {\widetilde{M}_T^2 \over 2}\, \s_1^2 - \widetilde{\mu}_T  \, \s_1^3 + \frac{1}{4}\lambda_T\, \s_1^4 +
 \lambda_1 \, \s_1^4 \, \log {\s_1^2\over B \, T^2} \,   .
 \ee
 Since this general form may originate from a large class of models beyond the 
specific case of Majoron models, the results here are of interest for a broad variety of applications.

In the Majoron model, the following identification holds:
\be
\lambda_1 \equiv  (N_{N_1} - 1) \,  {M^4 \, \over 32 \pi^2 \, v_0^4 } \,  , 
\;\;\; B \equiv {a_F \, v_0^2 \over M^2} \,  .
\ee

We obtain the
dimensionless effective thermal potential,
\be
\overline{V}^T_{\rm eff}(\overline{\s}_1,a_0,a_1) \equiv 4\,{\widetilde{\mu}_T^2 \over \widetilde{M}^6}\, V^T_{\rm eff}(\overline{\s}_1,\widetilde{\mu},\lambda_1) = 
{1 \over 2}\,\overline{\s}_1^2 - {1 \over 2}\,\overline{\s}_1^3 + {a_0 \over 8}\,\overline{\s}_1^4 + {a_1 \over 8}\,\overline{\s}_1^4 \, \log \overline{\s}_1^2 \,  ,
 \label{gen}
\ee 
which depends now on the two parameters,
\be
a_0 \equiv \widetilde{a} + a_1\, \log {\widetilde{M}_T^4 \over 4\, \widetilde{\mu}_T^2 \, B \, T^2} \
\ee
and
\be
a_1 \equiv {\lambda_1 \, \widetilde{M}_T^2 \over 2\, \widetilde{\mu}_T^2} \,  .
\ee

The Euclidean action that further generalizes Eqs.~(\ref{euclidean}) and (\ref{euclidean2}) can be written in the form,
\begin{eqnarray}\label{fa0a1}
\frac{S_3}{T} = \frac{\widetilde{M}_T^3}{\widetilde{\mu}_T^2 \, T}\, f(a_0,a_1) \,,
\end{eqnarray}
with
\be\label{fa0a1b}
f(a_0,a_1) = \pi \, \int_0^\infty \, d\bar{r}  \, \bar{r}^2 \left[ \frac{1}{2} \overline{\sigma}^2_1(\bar{r}) + \overline{V}_{\rm eff}^T(\overline{\s}_1, a_0, a_1) \right] \,,
\ee
where $\overline{\s}_1(\bar{r})$ is the solution of the equation of motion. 
In the limit $a_1 \ra 0$, we find $f(a_0,a_1) \ra f(\widetilde{a})$ and we recover Eq.~(\ref{euclidean2}). 
In Appendix~\ref{euclidapp} we provide an analytical expression for $f(a_0,a_1)$ by fitting numerical results.
%a fit we used to calculate the GW spectrum for the points generated in the scatter plot.

\subsection{GW signals}

Using the expressions for the Euclidean action in the previous subsection, for the different scenarios we calculate $\a$ and $\beta/H_{\star}$, and the resulting GW spectrum.
%scanning over the involved parameters.

\subsubsection{Minimal model}

\begin{figure}
\begin{center}
\psfig{file=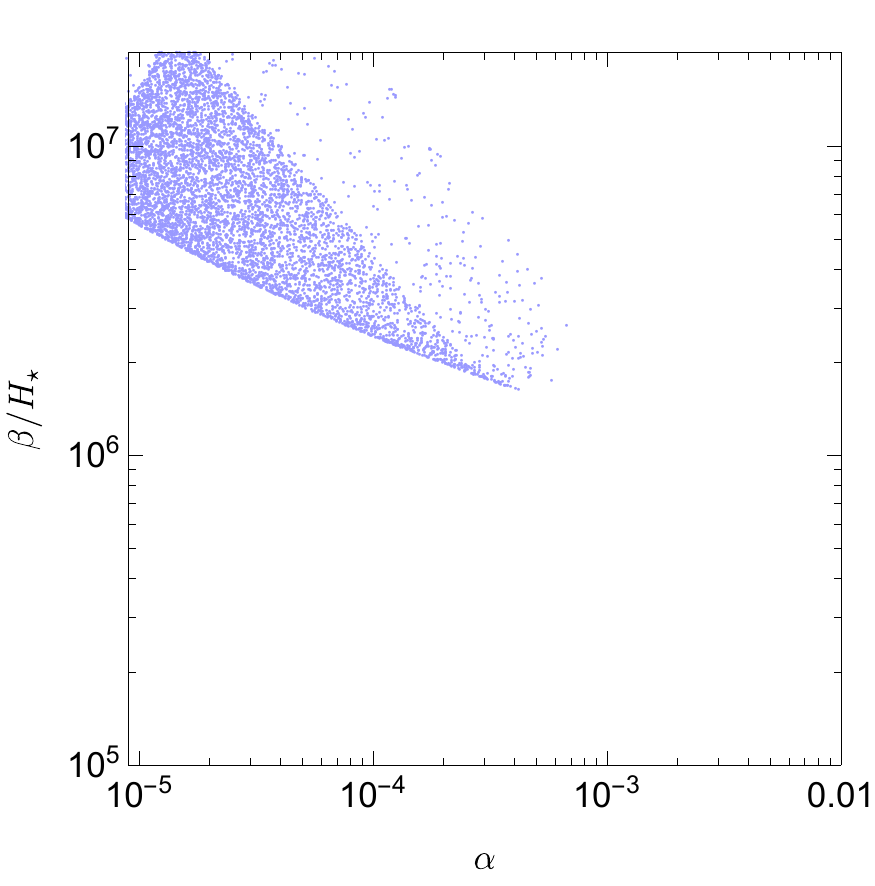,height=70mm,width=70mm} 
%\hspace{2mm}
%\psfig{file=f1b.pdf,height=58mm,width=64mm} 
\end{center}
%\vspace{-10mm}
\caption{Scatter plot in the $(\a, \beta/H_\star)$ plane for the minimal model.}
\label{f1}
\end{figure}

In this case we have three parameters, $v_0$,  $m_{\rm S}/v_0$,  $M/v_0$. The results for  $\a$ and $\beta/H_\star$ are shown in Fig.~\ref{f1} as a scatter plot. It can be seen that we have $\a \lesssim 10^{-3}$
and $\beta/H_\star \gtrsim 10^6$. The peak of the GW spectra are  six or seven orders of magnitude below the sensitivity
of any planned experiment. Therefore, we do not show GW spectra for this model.

\subsubsection{Adding an auxiliary real scalar}

Including an auxiliary real scalar field $\eta$ introduces a cubic term at zero temperature
and, as in the case of electroweak baryogenesis, may increase the strength of the phase transition 
and GW production. In addition to the three parameters of the minimal model, we
scan over $\widetilde{\mu}/v_0$, the parameter describing the zero-temperature cubic term. 
\begin{figure}
\begin{center}
\psfig{file=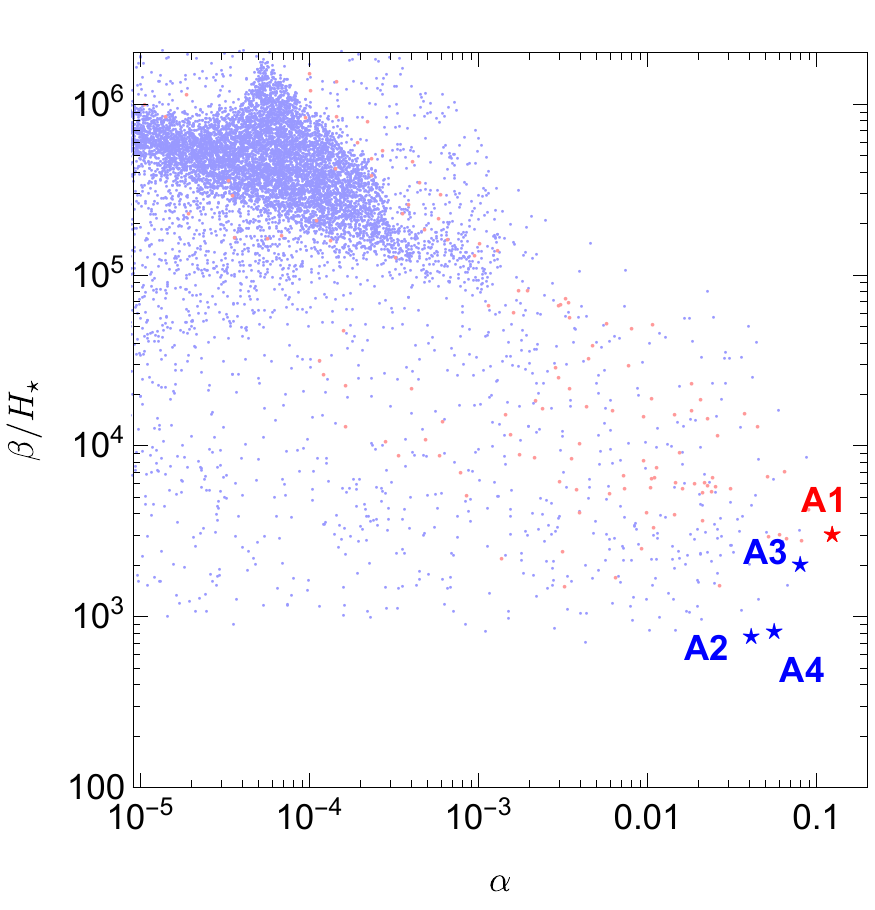,height=70mm,width=70mm} 
\hspace{2mm}
\psfig{file=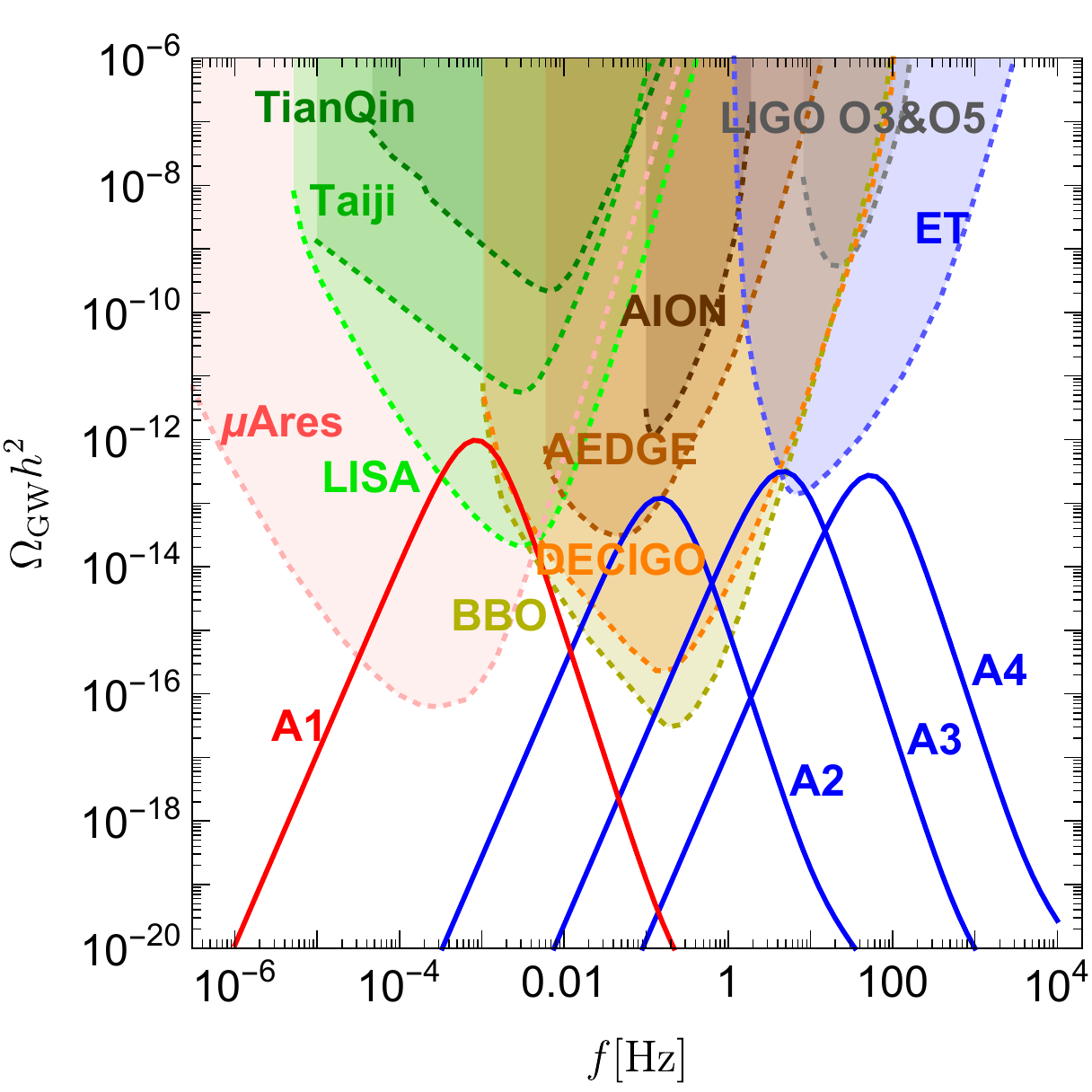,height=70mm,width=70mm} 
\end{center}
\caption{Quartic potential with zero temperature cubic term. Left panel: Scatter plot in the $(\a, \beta/H_\star)$ plane. The blue (red) points
correspond to the high scale (GeV seesaw scale) scenario. Right panel: GW spectra for the four benchmark points in Table~\ref{tab:samples_highT1} and marked with stars in the left panel.}
\label{f2}
\end{figure}
The result of the scan is shown as blue points in the left panel of Fig.~\ref{f2}.
 The points fall in the region with $\a \lesssim 0.1$ and $\beta/H_\star \gtrsim 1000$.
We select three benchmark points (A2, A3 and A4 marked with blue stars) whose parameters are in Table~\ref{tab:samples_highT1}. 
The corresponding GW spectra are displayed in the right panel; the scale of the phase transition covers many orders of magnitude, from 
$\sim$100 GeV to $\sim$PeV. The sensitivity of 
LIGO \cite{Aasi:2013wya,Abbott:2021xxi} and the future experiments, 
$\mu$Ares~\cite{Sesana:2019vho}, 
TianQin~\cite{Luo:2015ght}, 
Taiji~\cite{Guo:2018npi}, 
LISA~\cite{Auclair:2019wcv}, 
BBO~\cite{Yagi:2011wg}, 
DECIGO~\cite{Kawamura:2019jqt}, 
AEDGE~\cite{Bertoldi:2019tck}, 
AION~\cite{Badurina:2019hst}, 
ET~\cite{Hild:2010id},
are also shown. 
Clearly, the scenario is testable at 
AEDGE, DECIGO, BBO and  ET.\footnote{It may be possible to obtain even larger
 GW signals by engineering a supercooled phase transition in the conformal limit. However, this requires a significant extension of the Majoron models under consideration so that the scalar masses vanish at tree level and a false vacuum persists at zero temperature~\cite{Espinosa:2007qk}. 
This can be done by gauging $U(1)_L$ and introducing kinetic 
mixing with a $U(1)'$ in the dark sector as in the classically conformal $B-L$ model~\cite{Das:2016zue,Iso:2017uuu,Marzo:2018nov}.}
\begin{table}[t]
\begin{center}
%\begin{tabular}{ llllllll }
%\begin{table}[h]
%\begin{center}
\begin{tabular}{ l | cccc | cccc }
 \hline\hline
& \multicolumn{4}{c|}{Inputs} & \multicolumn{4}{c}{Predictions} \\
 \cline{2-9}
 &  $m_S/{\rm GeV}$ & $\tilde{\mu}/{\rm GeV}$ & $M/{\rm GeV}$ & $v_0/{\rm GeV}$ & $T_\star/{\rm GeV}$ & $\alpha$ & $\beta / H_\star$ & $\widetilde{a}$  \\\hline\hline
 A1 & $0.06190$ & $0.0005857$ & $0.5361$ & $3.5873$ & $0.6504$ & $0.1248$ & $2966$ & $0.05951$ \\
 \hline
A2 & $156.2$ & $13.15$ & $465.6$ & $1014$ & $721$ & $0.04139$ & $754.8$ & $0.3886$ \\
A3 & $1036$ & $13.72$ & $7977$ & $44424$ & $9180$ & $0.08012$ & $1975$ & $0.06268$ \\
A4 & $43874$ & $1856$ & $181099$ & $567378$ & $247807$ & $0.05611$ & $809.7$ & $0.1944$ \\
 \hline\hline
\end{tabular}
\end{center}
\caption{Values of the parameters corresponding to the four benchmark (starred) points in Fig.~\ref{f2}.}
%The first four correspond to the stars in Fig.~\ref{fig:scatterplot}, while the fifth is off scale and is not shown in Fig.~\ref{fig:scatterplot}.
\label{tab:samples_highT1}
\end{table}

\subsubsection{Nonthermal RH neutrinos}

We now consider the case with a nonthermal RH neutrino abundance which introduces 
a logarithmic term in the effective potential. We set
the zero-temperature cubic term $\widetilde{\mu} = 0$ so that the potential depends on the nonthermal RH neutrino abundance $N_{N_1}$
in addition to the three parameters of the minimal model. 
\begin{figure}[t]
\begin{center}
\psfig{file=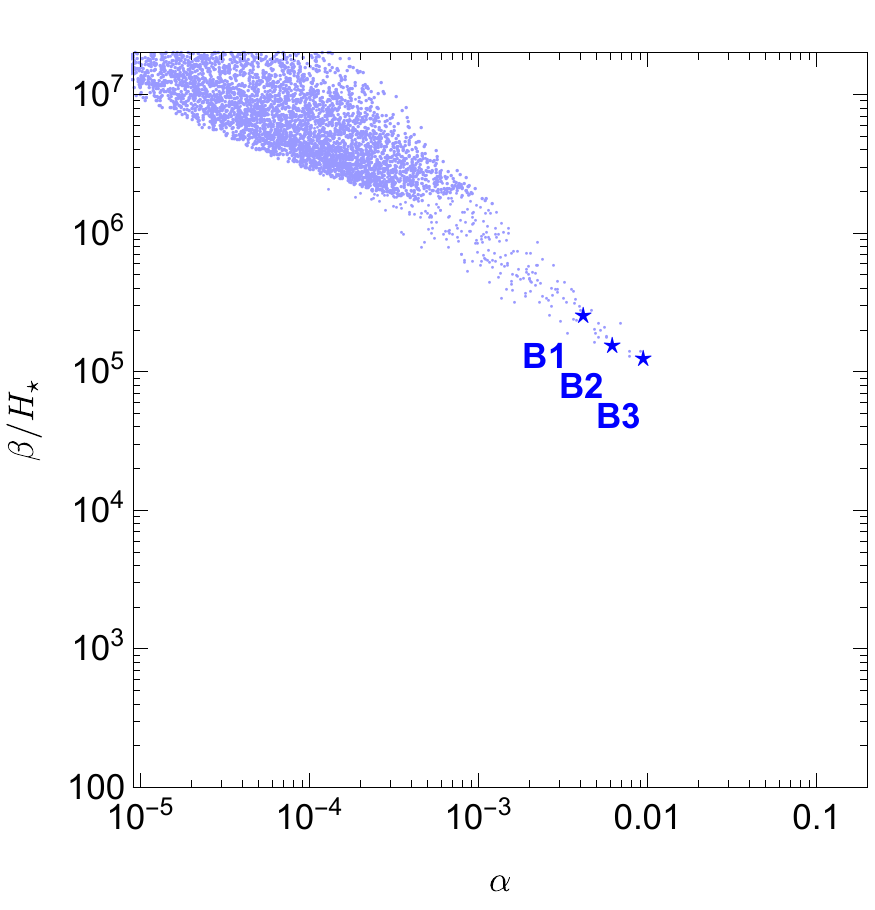,height=70mm,width=70mm} 
\hspace{2mm}
\psfig{file=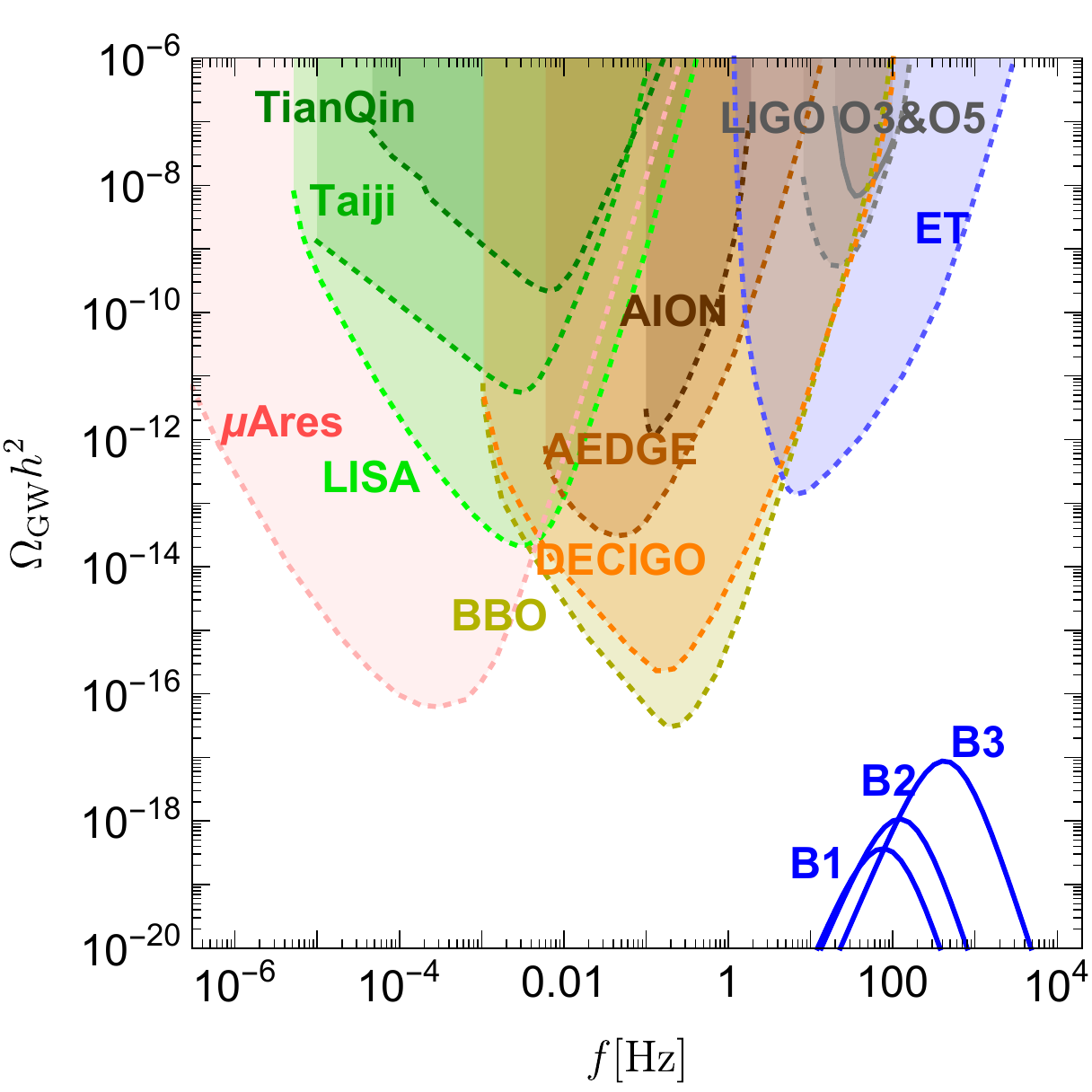,height=70mm,width=70mm} 
\end{center}
\caption{Potential with logarithmic field dependence and no cubic term at zero temperature. 
Left panel: Scatter plot in the $(\a, \beta/H_\star)$ plane. Right panel: GW spectra for the three benchmark points in Table~\ref{tab:samples_highT} and marked with stars in the left panel.}
\label{f3}
\end{figure}
The result of the scan is shown in the left panel of Fig.~\ref{f3}. We select three 
benchmark points (B1, B2 and B3 marked with stars) and shown the corresponding GW spectra in the right panel; see Table~\ref{tab:samples_highT} for the parameter values. Although the GW signals
are enhanced compared to the minimal model, they are several orders of magnitude below the experimental sensitivity.
\begin{table}[t]
\begin{center}
%\begin{tabular}{ llllllll }
%\begin{table}[h]
%\begin{center}
\begin{tabular}{ l | cccc | ccccc }
 \hline\hline
& \multicolumn{4}{c|}{Inputs} & \multicolumn{5}{c}{Predictions} \\
 \cline{2-10}
 &  $m_S\over {\rm GeV}$ & $N_{N_1}$ & $M\over {\rm GeV}$ & $v_0\over{\rm GeV}$ & $T_\star\over{\rm GeV}$ & $\alpha$ & $\beta / H_\star$ & $a_0$ & $a_1$  \\\hline\hline
 B1& $1383$ & $3.710$ & $1523$ & $1524$ & $991.4$ & $0.004190$ & $247858$ & $-0.04945$ & $4.314$ \\
 B2& $3225$ & $3.968$ & $3763$ & $3882$ & $2331$ & $0.006232$ & $152233$ & $2.898$ & $8.775$ \\
 B3& $19003$ & $7.108$ & $17724$ & $20947$ & $11682$ & $0.009441$ & $121398$ & $3.1422$ & $9.1367$ \\
 \hline\hline
\end{tabular}
\end{center}
\caption{Values of the parameters corresponding to the three benchmark (starred) points in the left panel of Fig.~\ref{f3} and with
GW spectra shown in the right panel of the same figure.}
\label{tab:samples_highT}
\end{table}

Our results are consistent with the general analysis presented
in Ref.~\cite{Ellis:2020awk} for potentials of the same form. However, in our case the range of $(\a, \beta/H_\star)$
in the scatter plot is not spanned because
the number of degrees of freedom in the dark sector is a small fraction of the SM value. Hence,
the GW signal is always weaker than, for example, from electroweak baryogenesis. 
The situation is better in low scale scenarios, but there is a price to pay.

\section{Low scale scenarios}
%%%%%%%%%%%%%%%%%

So far we have assumed that the phase transition in the dark sector occurs at a temperature above the electroweak scale, when all SM degrees of freedom are in ultrarelativistic thermal equilibrium. 
We also assumed that all  RH neutrinos acquire mass during the phase transition 
so that $T_\star$  approximately coincides with the seesaw scale, i.e., $T_\star \sim M$.  
In this section we relax either the first of these assumptions or both, thereby considering scenarios in which the  phase transition occurs at temperatures below the electroweak scale, and the seesaw scale $M$ does not necessarily coincide with $T_\star$.

\subsection{General considerations}

There is an important consequence of this new setup. In high scale scenarios, the Yukawa couplings of the seesaw RH neutrinos
couple the dark sector to the SM sector. During the phase transition 
the seesaw RH neutrinos acquire mass and then quickly decay. So the relic dark sector comprised of the two scalars $S$ and $J$ decouples after the phase transition and $T_{\rm dec} \simeq T_{\star}$.
This is why  we could take the dark sector and the SM thermal bath to have the same temperature during the phase transition,
and calculate the effective thermal potential at the common temperature $T$.
However, if the phase transition and seesaw scales are the same and lower than the electroweak scale, then from
Eq.~(\ref{Teq}) with $K_I \lesssim 500$ (to avoid fine-tuning),
the seesaw scale cannot be lower than $M \sim {\rm GeV}$ for $T^{\rm eq}_I \gtrsim 100\,{\rm GeV}$. 
Below this scale the seesaw RH neutrinos, and consequently the entire 
dark sector, will not thermalize and the maximum value of $\a$ sharply decreases, suppressing the GW signal.  

\subsection{GeV seesaw scale scenario}  

The possibility that the seesaw scale can be as low as $M \sim {\rm GeV}$ has been intensively investigated since it offers an avenue to test the seesaw mechanism directly at colliders and via meson decays, and also because such an extension of the SM does not destabilize
the electroweak scale, thus complying with naturalness. Moreover, for RH neutrinos in the GeV mass range, the matter-antimatter asymmetry of the universe can be explained by leptogenesis from RH neutrino oscillations~\cite{Akhmedov:1998qx}. 

We consider a scenario  with $T_{\rm dec} = T_{\star}\sim M \sim  {\rm GeV}$. 
%In this case one still has $r_T = 1$
Then, $g_{\rho}(T_\star \sim {\rm GeV}) = 61.75$ which increases the value of $\a$  (see Eq.~\ref{alpha}) compared to the
high scale scenarios.\footnote{A value $g_{\rho}(T_\star \sim {\rm GeV}) = 61.75$ holds
for a phase transition temperature just below half the tau mass, $T_\star \simeq 0.65 \, {\rm GeV}$.} 
The red points and benchmark point A1 in the left panel of Fig.~\ref{f2}  correspond to this scenario.  The parameter values for A1 and its GW spectrum are provided in Table~\ref{tab:samples_highT1} and the right panel of Fig.~\ref{f2}. 
Indeed, the GW signal has a higher peak than for the high scale benchmark points,
and the peak frequency lies within LISA's sensitivity range.\footnote{The proposed $\mu$Ares interferometer~\cite{Sesana:2019vho} has two orders of magnitude greater sensitivity than LISA in the mHz range, and will easily test the GeV scale scenario.} 
It is noteworthy that LISA is more sensitive to low scale phase transitions
with $\beta/H_\star \simeq 1000$,  than  electroweak scale phase transitions for which
$\beta/H_\star \simeq 10$--100 is typically assumed.

\subsection{Splitting the seesaw and the phase transition scales} 

As another class of low scale scenarios, we consider a phase transition scale much below a  GeV. 
For the time being, the two heaviest RH neutrinos, which we refer to as seesaw RH neutrinos,  
are assumed to couple the two sectors at temperature  $T \sim M \gtrsim 1\,{\rm GeV}$. (Later, we will 
relax this assumption and also consider other options.) Since $T_\star \ll M$, the two seesaw RH neutrinos  do not acquire  mass 
$\sim M$ during the $\s$-phase transition. Instead, they may gain their mass during an earlier phase transition at $T \sim M$,
as for example, in the case with an additional real scalar $\eta$.  While this will not play a role in our immediate considerations, it is interesting that a double peaked GW spectrum may result, with one peak at a high frequency and a second peak at a lower frequency. 
The $\s$-phase transition  will then give a mass only to the lightest  RH neutrino $N_1$, and, possibly, to additional lighter RH neutrinos. 
 We continue to denote the mass scale of the two heavy seesaw RH neutrinos by $M$. We will denote the 
 common mass of the light RH neutrinos generated during the $\sigma$-phase transition by $M'$,   and their number by $N' (=N-2)$. If there are no additional light RH neutrinos beyond $N_1$, then $N' = 1$.

For now we also assume that the $N'$ light RH neutrinos, and consequently the whole dark sector, does  not recouple  
prior to the phase transition, and therefore $T_{\rm dec} \simeq M \gg T_\star$. Then, at the phase transition the dark sector has 
a temperature $T' < T$. Our expressions for the thermal effective potential continue to hold
with the replacement $T \ra T'$. One then needs to express $T'$ as a function of $T$. This can be easily done by requiring entropy 
conservation between decoupling  and the phase transition. Introducing the entropy number degrees of freedom $g_s(T)$, 
the total entropy density is
\be
s(T) = {2\pi^2 \over 45} \,  g_s(T) \, T^3 \,  . 
\ee
Like the energy density, this can be split into a contribution from the SM sector and a contribution from  the dark sector. Since the dark sector
is at temperature $T'$ when the SM  sector is at temperature $T$,
$s(T) = s_{\rm SM}(T) + s_{\rm dark}(T'(T))$. Both contributions in turn can  
be expressed in terms of their respective entropy number of degrees of freedom via
\be
s_{\rm SM}(T) = {2\pi^2 \over 45} \, g_s^{\rm SM}(T) \,  T^3 \,  \;\;\;\;\; \mbox{\rm and} \;\;\;\;\;
s_{\rm dark}(T'(T)) = {2\pi^2 \over 45} \, g_{\rm dark}(T) \, T^3 \, r_T ^3 \,  ,
\ee
where  $r_T \equiv T'/T$ and $g_{\rm dark}(T) \equiv g_{\rm dark}(T'(T))$.  With this definition of $g_{\rm dark}(T)$, we can write
$g_s(T) = g_s^{\rm SM}(T) + g_{\rm dark}(T)\, r_T ^3 $.
We assume that when the two seesaw RH neutrinos decay, their entropy is redistributed in 
both sectors so that  $r_T(T_{\rm dec})= 1$.  In the dark sector, before the phase transition, $g_{\rm dark}(T)$ receives a contribution 
 from the scalars, $g_{\s+\eta} = 3$, and a contribution from the $N'$ light RH neutrinos, $g_{N'}= 7\,N'/4$. 
After decoupling, $g_{\rm dark}$ can be assumed to be constant until the phase transition.
Analogously to the standard calculation of the evolution of the ordinary neutrino temperature after decoupling, we find
\be\label{rT}
r_T(T_{\rm dec},T_\star) =  \left[ {g^{\rm SM}_s(T_\star) \over g^{\rm SM}_s(T_{\rm dec})} \right]^{1/3} \,  .
\ee
We further assume that after the phase transition, the massive scalar $S$ and the $N'$ light RH neutrinos decay
so that $g_{\rm dark}(T < T_\star) = 2$. In the next subsection we apply this general setup to a specific interesting scenario. 
  
\subsection{Phase transitions below an MeV, NANOGrav,  and the Hubble tension}  

An interesting possibility is whether the class of low scale scenarios with $T_\star \ll M$
can address the recent NANOGrav results which show evidence for the existence of a stochastic 
GW background at $\sim 10^{-8}\, {\rm Hz}$~\cite{Arzoumanian:2020vkk}. For $\beta/H_\star \gtrsim 1000$, 
Eq.~(\ref{fpeak}) indicates $T_{\star} \lesssim 100\, {\rm keV}$. The crucial question is if the amplitude
of the NANOGrav signal, $h^2\,\O_{\rm GW0}(f)\sim 10^{-10}$, is reproducible.
We will find that reproducing the NANOGrav signal is challenging~\cite{Arzoumanian:2021teu}. 
However, the predicted GW signal can be well within the reach of future experiments such as SKA~\cite{Janssen:2014dka} and THEIA~\cite{THEIA}. Quite interestingly, such a phase transition scale is independently motivated by  the Hubble tension, 
as we discuss below.

\begin{figure}
\begin{center}
\psfig{file=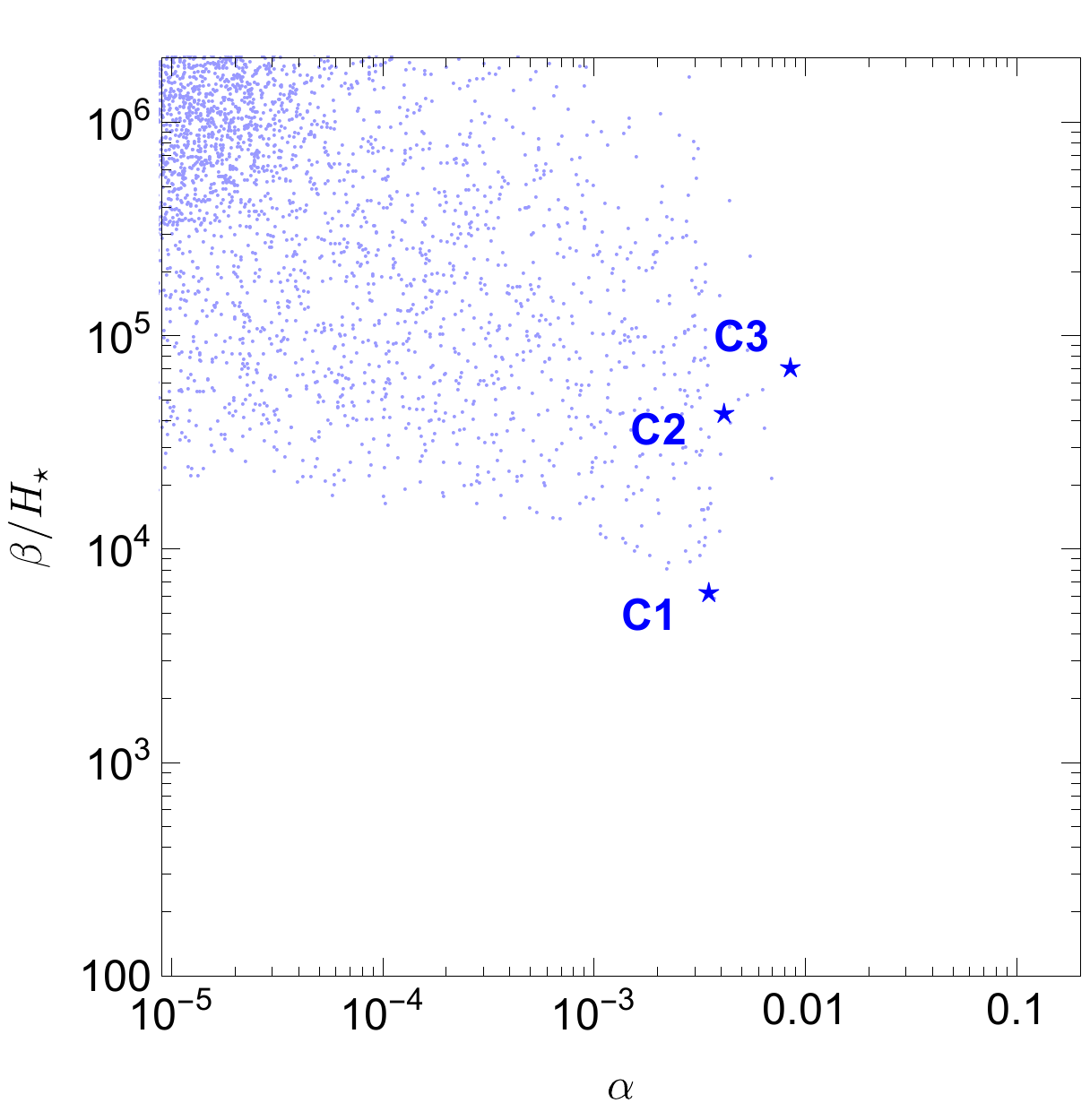,height=70mm,width=70mm} 
\hspace{2mm}
\psfig{file=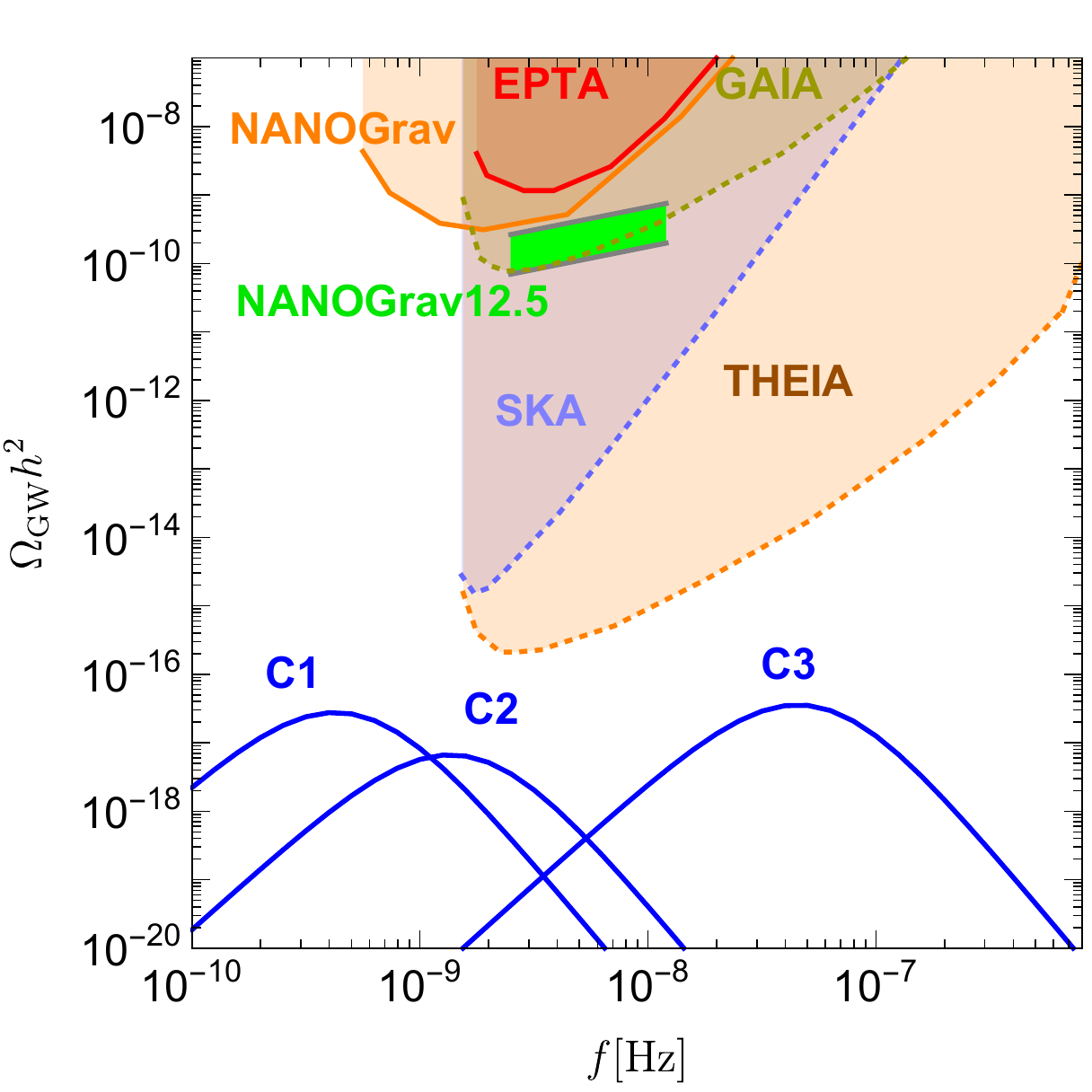,height=70mm,width=70mm} 
\end{center}
\caption{Low scale scenario for $T_{\rm dec} \gtrsim 100\,{\rm GeV}$
and $T_{\star} \lesssim 100\,{\rm keV}$. Left panel: Scatter plot in the $(\a, \beta/H_\star)$ plane.
 Right panel: GW spectra for the three benchmark points in Table~\ref{tab:samples_lowT1} and marked with stars in the left panel.}
\label{f4}
\end{figure}

For $T_\star \lesssim 100\,{\rm keV}$, well below the electron mass, the standard result
$g^{\rm SM}_s(T_\star) = 43/11$ holds and, since $g^{\rm SM}_s(T_{\rm dec} \gtrsim 100\,{\rm GeV}) = 106.75$,
 Eq.~(\ref{rT}) gives $r_T(T_{\rm dec} \gtrsim 100\,{\rm GeV},T_\star \lesssim 100\,{\rm keV}) \simeq 0.33$.
To calculate  $\a$ from Eq.~(\ref{alpha}), we also need 
\be
g_\rho^{\rm SM}(T_\star) = 2 + {7\over 4} \times 3.044 \times \left({4\over 11}\right)^{4\over 3} \simeq 3.36 \,,
\ee
 where  $N_{\nu}^{\rm eff, SM} \simeq 3.044$
is the effective number of ultrarelativistic neutrino species in the SM~\cite{Bennett:2020zkv,Akita:2020szl}.\footnote{The small 
deviation from $N_{\nu}^{\rm eff} = 3$ is due to the small component of nonthermal ordinary neutrinos produced in electron-positron annihilation.}
The left panel of Fig.~\ref{f4} shows for $N' =1$ 
that only $\a \lesssim 0.01$ and $\beta/H_\star \gtrsim 10^4$ are allowed. We select three benchmark points (C1, C2 and C3) that maximize the GW signal (shown in the right panel of Fig.~\ref{f4}; the parameter values are in Table~\ref{tab:samples_lowT1}).
 The signal peaks are many orders of magnitude
below the NANOGrav result and even below the sensitivity of future experiments like SKA and THEIA.
\begin{table}[t]
\begin{center}
%\begin{tabular}{ llllllll }
%\begin{table}[h]
%\begin{center}
\begin{tabular}{ l | cccc | cccc }
 \hline\hline
& \multicolumn{4}{c|}{Inputs} & \multicolumn{4}{c}{Predictions} \\
 \cline{2-9}
 &  $m_S/{\rm keV}$ & $\tilde{\mu}/{\rm keV}$ & $M/{\rm keV}$ & $v_0/{\rm keV}$ & $T_\star/{\rm keV}$ & $\alpha$ & $\beta / H_\star$ & $\widetilde{a}$  \\\hline\hline
C1 & $0.08266$ & $0.00002456$ & $0.04838$ & $0.09244$ & $0.3949$ & $0.003506$ & $6206$ & $0.5358$ \\
C2 & $0.01065$ & $0.000303$ & $0.03155$ & $0.02364$ & $0.1833$ & $0.004152$ & $42997$ & $0.7945$ \\
C3 & $0.5266$ & $0.2216$ & $0.8440$ & $0.2836$ & $3.831$ & $0.008528$ & $70771$ & $0.9583$ \\
 \hline\hline
\end{tabular}
\end{center}
\caption{Values of the parameters corresponding to the three benchmark (starred) points in Fig.~\ref{f4}.}
\label{tab:samples_lowT1}
\end{table}

Consider the possibility that the NANOGrav signal can be reproduced if the seesaw scale is lowered to a GeV, which as we have seen, permits thermalization of the SM and dark sectors. We set $T_{\rm dec}\sim 1\,{\rm GeV}$ and obtain
$r_T(T_{\rm dec} \sim 1\,{\rm GeV},T_\star \lesssim {\rm 100 keV}) \simeq 0.40$. This increases $\a$
with respect to the previous case (with $T_{\rm dec} \gtrsim 100\,{\rm GeV}$) by a factor 
$\sim (106.75/61.75)^{4/3} \sim (0.40/0.33)^{4} \simeq 2$. Since the peak of the GW spectrum
$\sim \a^5$ (accounting for the fact that $\beta/H_\star\propto \a^{-2}$~\cite{Ellis:2020awk})
this increases the signal by a factor  $\sim 38$, which is clearly insufficient to explain the NANOGrav signal and 
one can only marginally obtain testable signals at planned experiments.

Yet another possibility follows from the assumption that some additional interactions rethermalize the dark sector
with photons. Then, $r_T = 1$ which enhances the peak of the GW spectrum  by a factor of $\sim 10^9$. Although this is sufficient to reproduce the NANOGrav signal, one must contend with cosmological constraints, as we discuss below.

We may also allow the number of light RH neutrinos $N'$ to be larger than unity. This is expected
to increase the GW signal since these neutrinos contribute to the thermal effective potential; see Eqs.~(\ref{DA}) and~(\ref{lambdaT})
with $N$ replaced by $N'$. From this point of view $N'$ may be regarded as an effective
number of light RH neutrinos that represent a higher complexity of the dark sector. 

\subsubsection{Cosmological constraints on dark radiation}

Before proceeding, we pause to consider cosmological constraints on the extra radiation from the epochs of Big Bang Nucleosynthesis (BBN) and recombination. At temperatures $T \ll M$, the extra degrees of freedom are the $N'$ light RH neutrinos and the scalars $\eta, S$ and $J$.
The number of extra degrees of freedom at temperatures much below the muon mass is traditionally expressed
in terms of the effective number of extra neutrino species  $\D N_\nu^{\rm eff}(T)$ via
\be
g_\rho(T)  =  g_{\rho}^{\rm SM}(T) + {7 \over 4}\, \D N_\nu^{\rm eff}(T) \, [r_{T}^{\nu}(T)]^4 \,  ,
\ee
where 
\be\label{rnuT}
r_{T}^{\nu}(T) = \left({2 \over 11}\right)^{1\over 3}\, [g_s^{\g + e^{\pm}}(T)]^{1\over 3} \,  ,
\ee
with
\be
g_s^{\g + e^{\pm}}(T) =  2 + {45 \over \pi^4}\, \int_0^\infty \, dx \, {x^2 \,\sqrt{x^2 + z^2} +{1\over 3}\,{x^2 \over \sqrt{x^2 + z^2}} \over e^{\sqrt{x^2 + z^2}} + 1} \,  .
\ee
Here, $z \equiv m_{\rm e}/T$.
Primordial helium-4 abundance measurements combined with the baryon abundance extracted from Cosmic Microwave Background (CMB) anisotropies place a constraint on $\D N_\nu^{\rm eff}(t)$ at $t=t_{\rm f} \sim 1\,{\rm s}$,
the time of freeze-out of the neutron-to-proton ratio~\cite{Fields:2019pfx}:
\be\label{ubDNeetf}
\D N_\nu^{\rm eff}(t_{\rm f}) \simeq  -0.1 \pm 0.3 \ \Rightarrow \  \D N_\nu^{\rm eff}(t_{\rm f}) \lesssim 0.5 \, \ \  (95\% \, {\rm C.L.})\,  .
\ee
Measurements of the primordial deuterium abundance place a constraint on $\D N_\nu^{\rm eff}(t)$
at the time of nucleosynthesis, $t_{\rm nuc} \simeq 310 \, {\rm s}$, corresponding to $T_{\rm nuc} \simeq 65\,{\rm keV}$~\cite{DiBari:2018vba}:
\be\label{ubDNeetnuc}
\D N_\nu^{\rm eff}(t_{\rm nuc}) =  -0.2 \pm 0.3 \  \Rightarrow \  \D N_\nu^{\rm eff}(t_{\rm nuc}) \lesssim 0.4 \, \ \ (95\% \, {\rm C.L.}) \,  .
\ee
CMB temperature and polarization anisotropies constrain $\D N_\nu^{\rm eff}(t)$ at recombination, when 
$T \simeq T_{\rm rec} \simeq 0.3 \, {\rm eV}$,  and the {\em Planck} collaboration finds~\cite{Aghanim:2018eyx}
\be
\D N_\nu^{\rm eff}(t_{\rm rec}) =  -0.06 \pm 0.17\  \Rightarrow \  \D N_\nu^{\rm eff}(t_{\rm rec}) \lesssim 0.3 \, \ \ (95\% \, {\rm C.L.}) \,  .
\ee

We now calculate $\D N_\nu^{\rm eff}(t)$ for the Majoron model with the addition of a real scalar $\eta$.
Assuming that the dark sector decouples at a temperature $T _{\rm dec} \gtrsim 1\,{\rm GeV}$ and does not
recouple to the SM sector, the value of  $\D N_{\nu}^{\rm eff}$ is constant throughout BBN until recombination.
We evaluate  $\D N_{\nu}^{\rm eff}$ below $T \sim m_e/2 \simeq 250 \, {\rm keV}$
so that $r_{T}^{\nu}(T) \simeq (4/11)^{1/3}$, and obtain
%\footnote{Notice that here we have assumed that the massive scalars $S$ quickly decay in to the lighter RH neutrinos.}
\be\label{DNeff}
\D N_\nu^{\rm eff}(t_\star) = {4 \over 7}\,\left({11\over 4}\right)^{4\over 3}  \, r_T^4 \,g_{\rm dark}(T) \,  ,
\ee
with $g_{\rm dark}(T) =  3 + {7\over 4}\, N'$.
 %This is because the dark sector gets decoupled at temperatures much higher than $50\,{\rm MeV}$ and, therefore,
%$\D N_\nu^{\rm eff}$ can be taken as constant.  
%%%%%%%%%%%%%%%%%%%%%%%%%%%%%%%%%%%%%
In the canonical case with $T_{\rm dec} \gtrsim 100\,{\rm GeV}$ and $N'=1$, we have $r_T =0.33$ and 
$\D N_\nu^{\rm eff}(t_\star) \simeq 0.12$. 
In the low scale seesaw scenario with $T_{\rm dec} \simeq 1\,{\rm GeV}$ and $N'=1$, we have $r_T = 0. 4$ and
$\D N_\nu^{\rm eff}(t_\star) \simeq 0.27$.  These results are easily compatible with the BBN constraints.
After the phase transition and prior to recombination, the lightest RH neutrinos and scalar $S$ acquire mass and behave as matter before decaying. Because these particles redshift like matter, the later they decay, the greater the contribution to $\D N_\nu^{\rm eff}$ at recombination. Thus, their lifetimes must be bounded from above.  

 If some new interactions fully rethermalize the dark
sector prior to the phase transition, i.e., $r_T =1$, then for $N' = 1$,
$\D N_\nu^{\rm eff}(T_\star) \simeq 10$. This result could be put in agreement with BBN constraints
if the late thermalization occurs just prior to the phase transition and 
$T_\star \lesssim 65 \, {\rm keV}$, but it would be in obvious disagreement 
with the CMB anisotropies. 
There is, however, another interesting possibility to be considered.

\subsubsection{GW signals and the Hubble tension}

\begin{figure}[t]
\begin{center}
\psfig{file=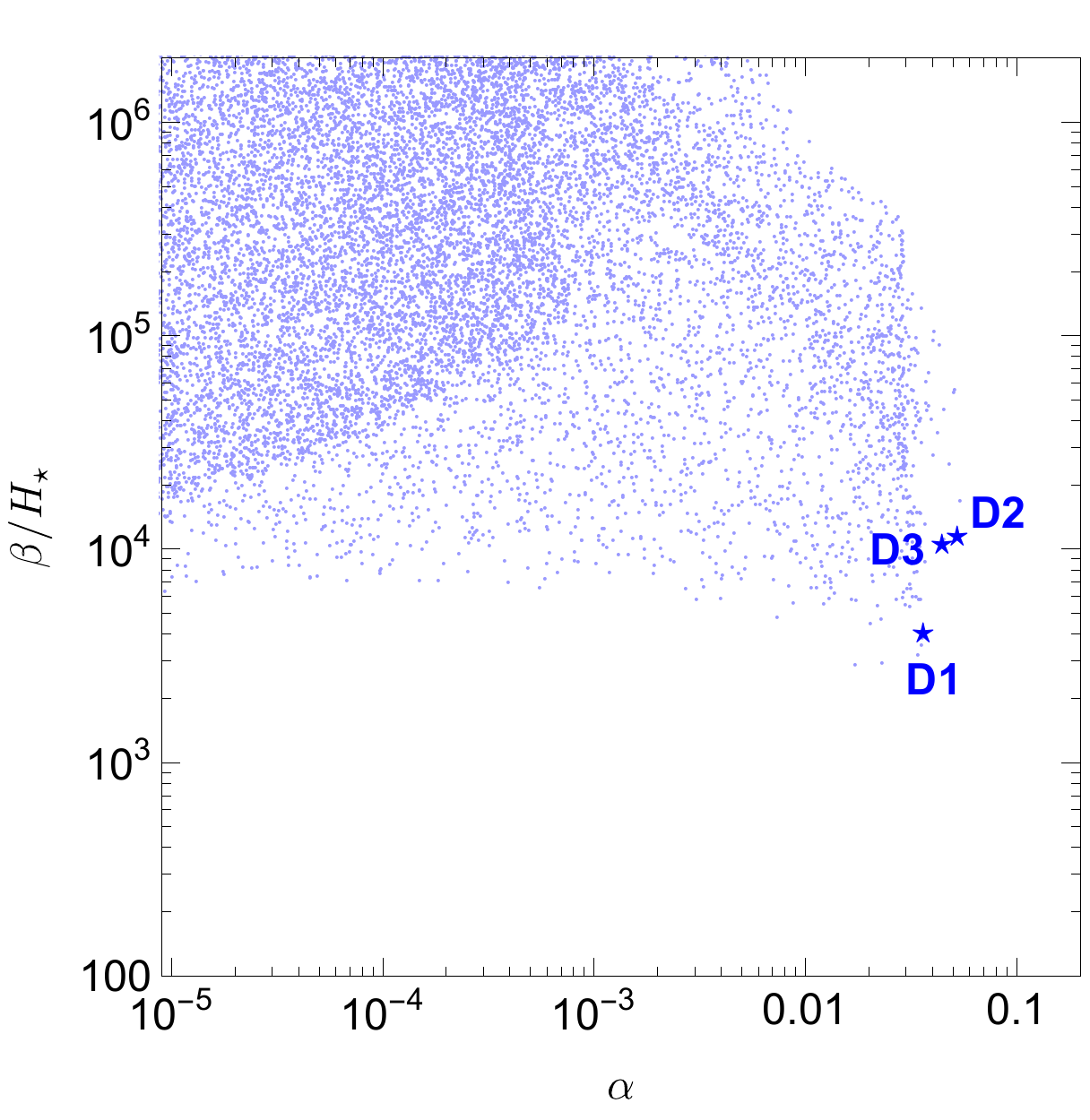,height=70mm,width=70mm} 
\hspace{2mm}
\psfig{file=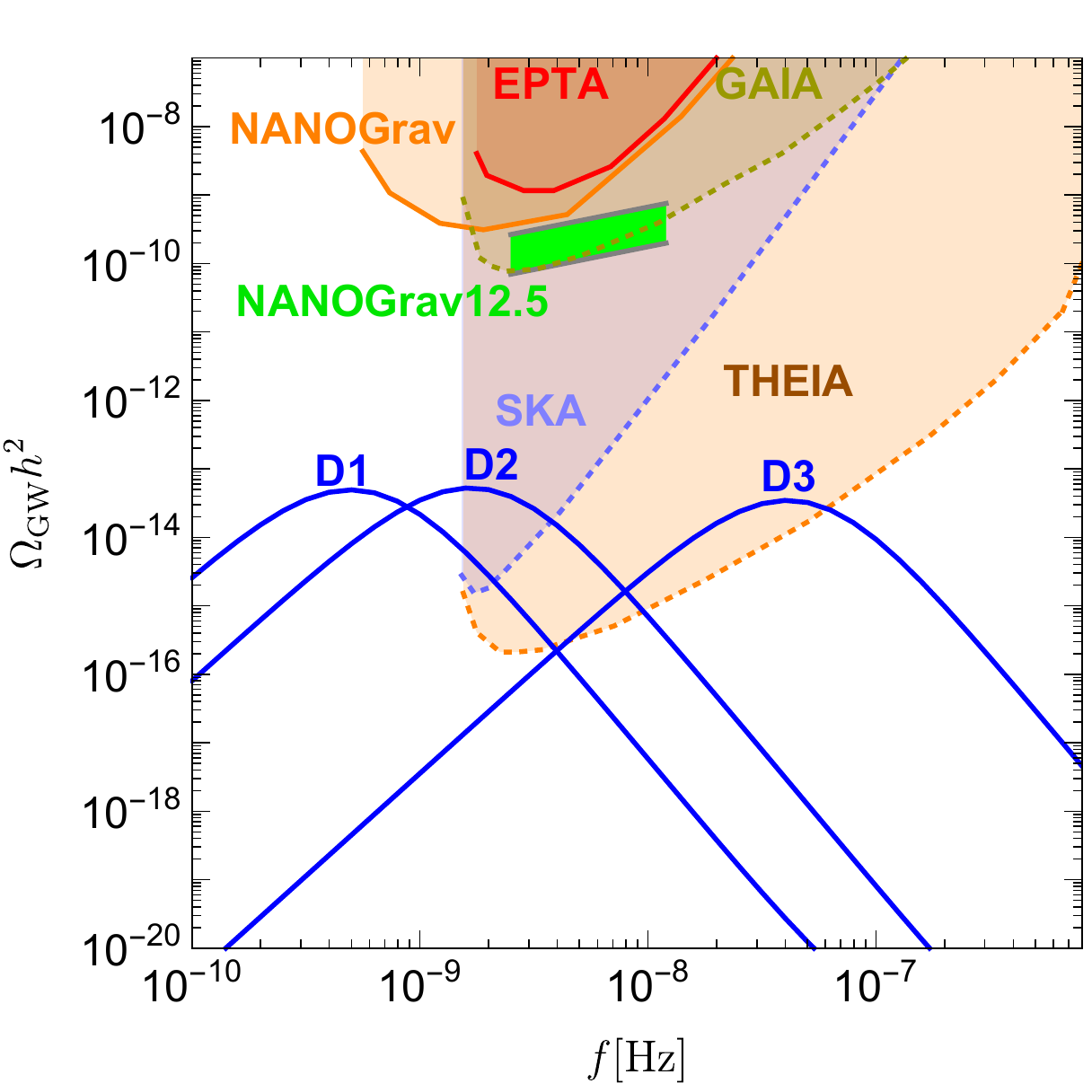,height=70mm,width=70mm} 
\end{center}
\caption{Rethermalization scenario with $r_T = 0.6$ and $N' =1$. Left panel: Scatter plot in the $(\a, \beta/H_\star)$ plane. Right panel: GW spectra for the three benchmark points in Table~\ref{tab:samples_lowT2} and marked with stars in the left panel.}
\label{f5}
\end{figure}

The so-called Hubble tension~\cite{Bernal:2016gxb}
between the CMB measurement of $H_0$ within $\L$CDM and low-redshift astrophysical measurements suggest  $\D N_\nu^{\rm eff}(t_{\rm rec}) \simeq 0.6$. %A model that relies on the decay of a long-lived massive particle into dark radiation to give
%This was at least initially the case with first results from {\em Planck} satellite  in 2013 \cite{Ade:2013zuv}. 
 %such a $\D N_\nu^{\rm eff}$ value  at recombination is presented in Ref.~\cite{DiBari:2013dna}. 
We calculate the maximum value of $r_T$ by very conservatively requiring $\D N_\nu^{\rm eff}(T_\star) \lesssim 1.35$ to address the Hubble tension. From Eq.~(\ref{DNeff}) with $N'=1$, we find
 $r_T \simeq 0.6$. The result of our scan  with $r_T = 0.6$ is shown in the left panel of Fig.~\ref{f5}. Higher values of $\a$ and lower values
 of $\beta/H_{\star}$ can be obtained compared to the case with $r_T =0.33$ presented in Fig.~\ref{f4}. 
 From the right panel we see that the GW spectra for three benchmark
 points can be detected at SKA and THEIA although their peaks are still three orders of magnitude below
 the NANOGrav signal.
 \begin{table}[t]
\begin{center}
%\begin{tabular}{ llllllll }
%\begin{table}[h]
%\begin{center}
\begin{tabular}{ l | cccc | cccc }
 \hline\hline
& \multicolumn{4}{c|}{Inputs} & \multicolumn{4}{c}{Predictions} \\
 \cline{2-9}
 &  $m_S/{\rm keV}$ & $\tilde{\mu}/{\rm keV}$ & $M/{\rm keV}$ & $v_0/{\rm keV}$ & $T_\star/{\rm keV}$ & $\alpha$ & $\beta / H_\star$ & $\widetilde{a}$  \\\hline\hline
D1 & $0.083651$ & $0.009534$ & $0.3574$ & $0.4082$ & $0.8076$ & $0.03641$ & $4046$ & $0.6795$ \\
D2 & $0.2650$ & $0.9064$ & $0.5571$ & $0.2505$ & $0.9902$ & $0.05283$ & $11407$ & $0.9207$ \\
D3 & $5.4732$ & $0.9724$ & $14.19$ & $8.341$ & $26.13$ & $0.04452$ & $10472$ & $0.8831$ \\
 \hline\hline
\end{tabular}
\end{center}
\caption{Values of the parameters corresponding to the three benchmark (starred) points in Fig.~\ref{f5}.}
\label{tab:samples_lowT2}
\end{table}

This scenario, however, is not only too generic, since we have not specified the interactions that cause
rethermalization, but also unrealistic, since we have assumed that the dark sector essentially equilibrates with photons.
Also, CMB data now clearly show that a mere injection of extra radiation at the level of
 $\D N_\nu^{\rm eff} \simeq 0.6$, though able to yield a higher value of $H_0$ from CMB anisotropies in agreement with
the astrophysical measurements, results in a worse global fit of the data. This is because the 
extra radiation also produces a shift in the acoustic peaks which is
not tolerated by current  CMB data~\cite{Knox:2019rjx}. 
To ameliorate the Hubble tension within an extension of $\Lambda$CDM {\it and} provide a better global fit to the data
requires a reduction of the sound horizon at recombination without an alteration of other CMB observables that are well fitted. 

It is quite interesting that the requirements of specifying the interaction within a realistic model 
and producing a better cosmological global fit can be neatly satisfied  within our Majoron model with split seesaw and phase transition scales. We introduce an interaction of the Majoron $J$ with the ordinary neutrino cosmic background of the type,
\be\label{nudark}
-{\cal L}_{\rm \nu-{\rm dark}} = {i\over 2}\,\sum_{i=2,3}\lambda_i \, \overline{\nu_{i}} \, \g^5 \,\nu_{i} \, \eta + 
{i\over 2}\,\lambda_1 \overline{\nu_{1}}\,\g^5 \,\nu_{1} \, J  + {\rm h.c.}\,  ,
\ee
where $\nu_{2,3}$ are the two
heavier ordinary neutrinos with masses generated  by the seesaw and $\nu_1$ is the lightest ordinary neutrino that couples to the Majoron.
Indeed, the interaction between dark radiation and neutrinos is just the ingredient needed to reduce the sound horizon and obtain
a larger value of $H_0$ from the CMB anisotropies.
The impact of this kind of interaction on the CMB was first studied  in Ref.~\cite{Chacko:2003dt} and it has been recently considered in 
Refs.~\cite{Escudero:2019gvw,Escudero:2021rfi} in the context of the Hubble tension. We have introduced a flavor dependence since in our case the seesaw scale and the phase transition scale are different. However, from the point of view
of the CMB anisotropies the flavor dependence has no effect. (This kind of model falls into a more general class of  models
recently discussed in Ref.~\cite{Blinov:2020hmc}, also in relation to the Hubble tension. A recent example of a specific model within this class, though  not a Majoron model, has been recently discussed in Ref.~\cite{Choi:2020pyy}.) The interactions in Eq.~(\ref{nudark})
 partly rethermalizes the dark sector with the ordinary neutrinos such that
\be
r_T =  \left({4 \over 11}\right)^{1\over 3}\, \left({3.044 \over 3.044 + N' + 12/7} \right)^{1 \over 4} \,  .
\ee
For  $N' =1 $, we again find $r_T \simeq 0.6$. Also,
\be\label{extra}
\Delta N_\nu^{\rm eff}(t_{\rm rec}) \simeq 3.044 \, \left({3.044 + N' + 12/7 \over 3.044 + N' + 12/7 - N_{\rm h}} \right)^{1\over 3} - 3.044\simeq 0.5 \,  ,
\ee
where $N_{\rm h}$ (equal to 2 in our case) is the number of massive states that decay and produce the excess radiation.
The value, $\Delta N_\nu^{\rm eff} \simeq 0.5$,  nicely coincides with the 
best fit value found in Refs.~\cite{Escudero:2019gvw,Blinov:2020hmc} with
$\Delta \chi_{\rm min}^2 = -12.2$ compared to the $\Lambda$CDM model; the fit
includes the astrophysical determination of $H_0$.\footnote{$\Lambda$CDM gives $\chi_{\rm min}^2 = 
2786.7$ \cite{Escudero:2019gvw}.} The best fit values for the coupling constant and Majoron mass  are $\lambda_{i} \sim 10^{-12}$ and  $m_J \sim {\cal O}(0.1 \,{\rm eV})$.  It is quite intriguing that such values for the Majoron mass can be accommodated in phase transitions with $T_\star \sim {\rm keV}$. These values give rise to GW signals in the $10^{-8}\,{\rm Hz}$ range tested by NANOGrav, SKA and THEIA.\footnote{It is conceivable that $m_J \sim {\cal O}(0.1)\,{\rm eV}$ may not generate a linear term in the potential which spoils the phase transition but a dedicated analysis would be needed. However,  note also that in our model, interactions involving the light RH neutrinos such as $\bar{\nu}_1\,\gamma^5 \, N_1 \, J $ and $i\, \bar{N_1} \gamma^5 \, N_1 \, J$ could be included. These will likely enlarge the parameter space of preferred values of $\lambda_i$ and $m_J$ that resolve the Hubble tension.}

 %%%%%%%%%%%
 The interactions in Eq.~(\ref{nudark}) now provide a realistic way to achieve rethermalization of the dark sector and justifies $r_T \simeq 0.6$ which links possible GW signals at $\sim 10^{-8}\,{\rm Hz}$ to the Hubble tension. Effectively, the interaction causes ordinary neutrinos to be treated as part of the dark sector rather than the SM sector.
 Compared to the case in which the dark sector thermalizes with photons, now the thermal coupling 
 between the dark sector and ordinary neutrinos also affects the value of $g_{\rho}^{\rm SM}(T_\star)$, and therefore the value of $\a$.
  This is because the neutrino-to-photon temperature ratio decreases from the standard value 
 $r^\nu_{T } = T_{\nu}/T = (4/11)^{1\over 3}$ to $r_T$. The total number of SM ultrarelativistic degrees of freedom at the phase transition  decreases from the standard value $g_{\rho }^{\rm SM} \simeq 3.36$ to
 \be
 g_\rho^{\rm SM'}(T_\star \lesssim 100\,{\rm keV}) = 2 + 3.044 \, {7\over 4} \, r_T^4  \, .
 \ee
For $r_T = 0.6$, $g_\rho^{\rm SM'}(T_\star \lesssim 100\,{\rm keV}) \simeq 2.69$. 
On the other hand, the contribution from the dark sector is 
\be
g_\rho^{\rm dark}(T_\star \lesssim 100\,{\rm keV}) = {7 \over 4}\, \left({4\over 11}\right)^{4\over 3}\, \Delta N_{\nu}^{\rm eff}(t_\star)\, \,  ,
\ee
where  $\Delta N_{\nu}^{\rm eff}(t_\star)$ is given by Eq.~(\ref{DNeff}).\footnote{Note that
$g_\rho^{\rm SM'} + g_\rho^{\rm dark} = g_{\rho}^{\rm SM}$. This is expected since 
thermalization transfers energy from ordinary neutrinos
to the dark sector with no change in the total energy density. The extra radiation described by Eq.~(\ref{extra}) is 
produced by decays of the massive states after the phase transition.}

\begin{figure}
\begin{center}
\psfig{file=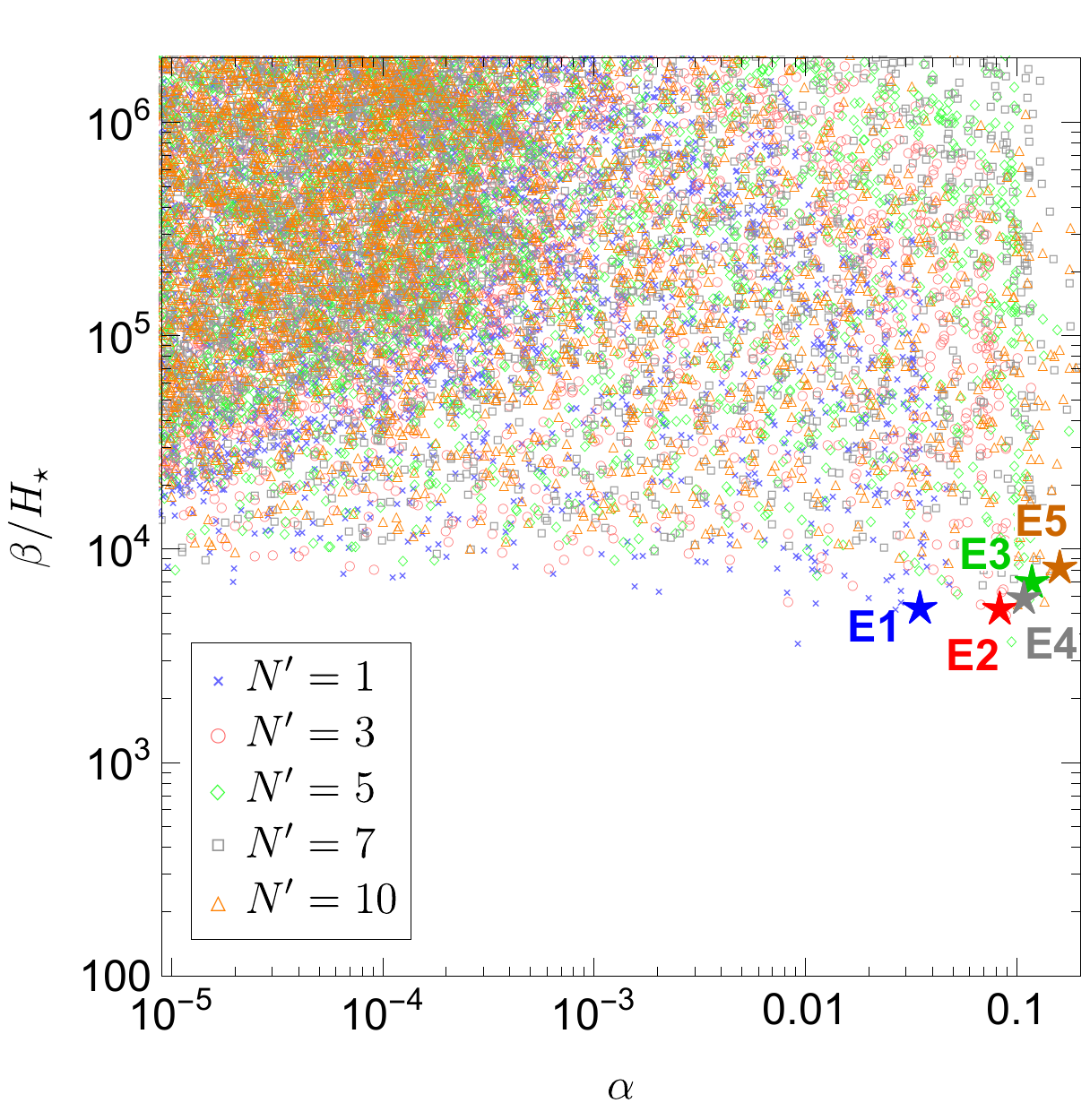,height=70mm,width=70mm} 
\hspace{2mm}
\psfig{file=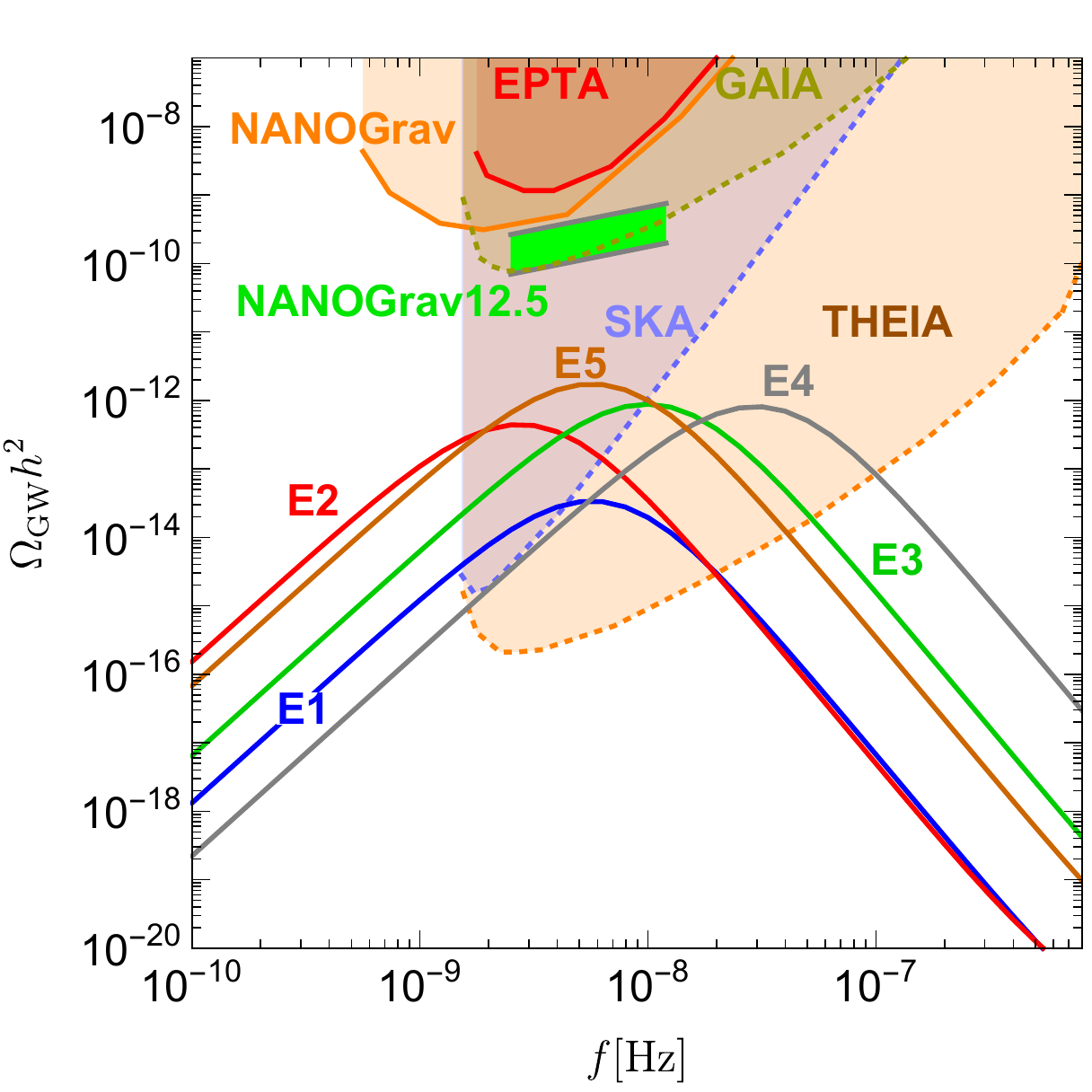,height=70mm,width=70mm} 
\end{center}
\caption{Rethermalization scenario addressing the Hubble tension.  Left panel: Scatter plot in the $(\a, \beta/H_\star)$ plane.  
 Right panel: GW spectra for the five benchmark points E1--E5 in Table~\ref{tab:samples_lowT3} corresponding to $N' =1 ,3, 5, 7, 10$, respectively, and marked with stars in the left panel. }
\label{f6}
\end{figure}

We have so far considered the case $N' = 1$ corresponding to the usually considered total number of RH neutrinos $N=3$. However, we can also study how the GW signal  changes with increasing $N'$. In this context, $N'$ should be regarded as an effective number
of light RH neutrinos parameterizing an extension of the dark sector that contributes to the effective thermal potential.
The result, not surprisingly, is that the GW signal is enhanced by increasing $N'$, but the enhancement saturates for $N' \sim 10$
as can be seen from the right panel of Fig.~\ref{f6}. We have selected five benchmark points with $N'= 1,3,5,7, 10$, and whose GW spectra peak in the range of frequencies
tested by NANOGrav, SKA and THEIA. The reason for the saturation is that for a given choice of all the other parameters, there is 
an upper bound on the value of $N'$ imposed by the stability of the potential, since the quartic coupling $\lambda_T$ in Eq.~(\ref{lambdaT}) 
becomes negative for too large $N$. Note that for $N' = 10$, Eq.~(\ref{DNeff}) gives
$\D N_{\nu}^{\rm eff}(t_{\rm nuc}) = 0.535$, in conflict with the BBN constraint in Eq.~(\ref{ubDNeetnuc}).
However, this can be circumvented by relaxing our assumption that
the dark sector undergoes an early thermalization at temperatures above $\sim 100\,{\rm GeV}$ 
and then rethermalizes at low temperatures. Instead of rethermalization, the dark sector could thermalize for the
first time at low temperatures prior to the phase transition. 
\begin{table}[t]
\begin{center}
%\begin{tabular}{ llllllll }
%\begin{table}[h]
%\begin{center}
\begin{tabular}{ l | cccc | cccc }
 \hline\hline
& \multicolumn{4}{c|}{Inputs} & \multicolumn{4}{c}{Predictions} \\
 \cline{2-9}
 &  $m_S/{\rm keV}$ & $\tilde{\mu}/{\rm keV}$ & $M/{\rm keV}$ & $v_0/{\rm keV}$ & $T_\star/{\rm keV}$ & $\alpha$ & $\beta / H_\star$ & $\widetilde{a}$  \\\hline\hline
E1 & $0.6847$ & $0.04274$ & $2.795$ & $3.507$ & $7.036$ & $0.03505$ & $5286$ & $0.7226$ \\
E2 & $0.3065$ & $0.008314$ & $1.401$ & $3.253$ & $3.711$ & $0.08406$ & $5229$ & $0.4638$ \\
E3 & $0.7224$ & $0.01154$ & $3.808$ & $12.34$ & $10.33$ & $0.1186$ & $6993$ & $0.7279$ \\
E4 & $2.620$ & $0.05962$ & $10.27$ & $35.12$ & $36.26$ & $0.1086$ & $5919$ & $0.3974$ \\
E5 & $0.2771$ & $0.003530$ & $1.402$ & $7.101$ & $5.207$ & $0.1609$ & $8176$ & $0.2127$ \\
 \hline\hline
\end{tabular}
\end{center}
\caption{Values of the parameters corresponding to the five benchmark (starred) points in Fig.~\ref{f6}.
}
\label{tab:samples_lowT3}
\end{table}

These results are quite intriguing,  since they show how a solution to the Hubble tension motivates a GW signal from a strong first-order phase transition in Majoron models  within NANOGrav's range of frequencies. We also see that even increasing $N'$ yields GW spectra
with peaks two orders of magnitude below the NANOGrav signal. We do not expect that a proper accounting of theoretical uncertainties or a calculation of the signal-to-noise ratio can overcome the two orders of magnitude needed to explain the NANOGrav signal.

\section{Final remarks and conclusion}

We discussed how a stochastic GW background may be generated in the early universe from
first-order phase transitions in Majoron models. We considered high scale scenarios, in which the
phase transition occurs above the electroweak scale, and  low scale scenarios in which the phase transition occurs below the electroweak scale.
%In high scale scenarios the signal can be tested with future interferometers. 

We highlight some aspects of our study.
\begin{itemize}
\item We have not introduced couplings of the complex scalar $\s$
to the Higgs boson. So  the Higgs sector is not modified, for example by
a Higgs portal operator $\propto H^\dagger H \, |\s|^2$. In the high scale scenarios, with $v_0 \gg v_{\rm ew}$, it can be assumed
that the coupling is small  to not destabilize the electroweak scale and introduce a naturalness problem. 
Below the electroweak scale, this term would simply contribute to the quadratic term and be incorporated into $\mu^2$. We  have also not explicitly considered nonrenormalizable higher dimensional
operators so that our model is ultraviolet-complete. 
\item In high scale scenarios the GW signal can be detected at currently planned interferometers. 
\item The stability of the effective potential limits the size of the GW signal at low frequencies
in the low scale scenarios to two or more orders of magnitude below the NANOGrav signal. If the NANOGrav signal is confirmed
% and a spectral shape consistent with a phase transition be observed, then further investigation would be motivated, calculating the signal-to-noise ratio and reducing theoretical uncertainties.  If the such a result is confirmed, 
then a more drastic departure from the models presented here needs to be considered.
\item For phase transitions in the dark sector, typically $\beta/H_\star \gtrsim 1000$ rather than
$\beta/H_\star \gtrsim 10$--$100$, as usually assumed in the case of electroweak phase transitions. This shifts all signals to higher frequencies, which makes it desirable that experiments be planned with greater sensitivity in the range of frequencies relevant to ET. We find that LISA will 
test phase transitions at the GeV scale that may have connections with Majoron models, and with RH neutrino searches at FASER~\cite{Ariga:2019ufm}.
\item In low scale scenarios we considered the possibility that the phase transition occurs even below the decoupling scale of ordinary neutrinos. In this case, a GW signal is generated in the range of frequencies currently probed by NANOGrav and may be observable 
at SKA and THEIA. 
\item The Hubble tension can be addressed in models in which the dark sector interacts with the ordinary neutrinos. 
This offers an intriguing connection between two apparently independent phenomenologies and  makes it possible to test late thermalization scenarios and departures from $\L$CDM with GW experiments.

\end{itemize}

Finally, we stress the {\it inverse problem}, i.e.,
the possibility of extracting the physical parameters that describe the phase transition from an observed GW signal. 
The peak frequency, amplitude and shape of the spectrum can help reconstruct 
the four thermodynamic parameters, $\a,T_\star,\beta/H_\star,v_{\rm w}$, that describe the phase transition despite the presence of degeneracies.
 Recently, it has been emphasized that a more
realistic description of the sound wave contribution entails a double broken power law
parameterized by four spectral parameters rather than just two for a single power law as in Eq.~(\ref{omegasw})~\cite{Gowling:2021gcy}.
It should be said however that the situation for values of $\beta/H_\star \gtrsim 1000$,
 as in the case of a dark sector phase transition, is much more challenging than in the case of a phase transition
 in the visible sector such as the electroweak phase transition. For this reason it is important to have a specific model
in which the GW signal may be linked to other phenomenological observables.
Our study meets this requirement since 
 there is a strong complementarity with other phenomenology, such as meson decays experiments for
GeV scale scenarios, and cosmological tensions (i.e., departures from $\L$CDM) in the case of very low energy scale
scenarios. We have also seen how in the latter case a very interesting signature  is the possibility of observing a 
double peak GW signal, one at high frequencies from seesaw RH neutrinos and one at low frequencies from light RH neutrinos.

%Interestingly, LISA will be able to test GeV seesaw scale scenarios also associated to low scale leptogenesis and RH neutrino searches at experiments such as FASER. 

%\newpage
%\vspace{-1mm}
\section*{Acknowledgments} 

PDB and YLZ acknowledge financial support from the STFC Consolidated Grant ST/T000775/1. 
This project has received funding/support from the European Union Horizon 2020 research and innovation 
programme under the Marie Sk\l{}odowska-Curie grant agreements number 690575 and  674896. 
DM is supported in part by U.S. DOE Grant No. de-sc0010504.

\appendix
\renewcommand{\thesection}{\Alph{section}}
\renewcommand{\thesubsection}{\Alph{section}.\arabic{subsection}}
\def\theequation{\Alph{section}.\arabic{equation}}
\renewcommand{\thetable}{\arabic{table}}
\renewcommand{\thefigure}{\arabic{figure}}
\setcounter{section}{0}
\setcounter{equation}{0}

%\newpage
%%%%%%%%%%%%%%%%%%%%%%%%%%%%%%%%%%%%%%%%%%%%%
\section{Appendix}
%%%%%%%%%%%%%%%%%%%%%%%%%%%%%%%%%%%%%%%%%%%%%

In this Appendix we briefly discuss how to calculate $T_{\star}$ and derive
Eq.~(\ref{SETstar}). We also present analytical fits of the Euclidean action for a general effective potential.  All results contained in this Appendix are applicable in general,  and
beyond Majoron models.

\subsection{Calculation of $T_{\star}$}
\label{app1}
%-------------------------------------------

We defined the time of the phase transition by $I(t_\star) = 1$; see Eq.~(\ref{It}).  A brute force method of calculating $T_\star$
is to plug an expression for the
Euclidean action $S_E$ as a function of temperature into the equation, $I(t(T_\star))=1$, and solve it numerically. However,
Eq.~(\ref{It}) can be simplified because in the detonation regime the duration of the 
phase transition is much shorter than the age of the universe.  This is confirmed by the fact we always find $\beta/H_{\star} \gg 1$.  We can then neglect the expansion during the phase transition and write,
\be\label{Itb}
I(t) \simeq {4\pi \over 3} \,v_{\rm w}^3  \int_{t_{\rm c}}^t \, dt' \,  \Gamma(t') \, (t - t')^3 \,  .
\ee
Since we know the nucleation rate as a function of temperature,
\be
\Gamma(T) \simeq T^4 \, \left[{S_E(T)\over 2 \pi }\right]^{3/2} \, e^{-S_E(T)} \,  ,
\ee
it is convenient to switch variables from time to temperature using $dt' =  -dT' /[H(T')\,T']\, $ and the usual relation in the radiation-dominated regime,
\be
t = {1 \over 2}\,\sqrt{90 \over 8\pi^3 \, g_{\rho}(T)} \, {M_{\rm P} \over T^2} \,.
\ee
In terms of the dimensionless 
quantity $x \equiv T/v_0$, the condition $I(T_\star) = 1$ is equivalent to
\be\label{hv0}
h(x_\star) = C(v_0) \,  ,
\ee
where
\be
h(x) \equiv \int_{x_\star}^{x_{\rm c}}\,dx' \,  x' \, \left[{S_E(x') \over 2\,\pi}\right]^{3 \over 2} \, e^{-S_E(x')}
\,\left( {1\over x^2} - {1\over x^{'2}}\right)^3 \, 
\ee
and
\be
C(v_0) \equiv { H^4(v_0) \over v_0^4} \, {6 \over \pi \, v_{\rm w}^3} \,  .
\ee
Although laborious, solving Eq.~(\ref{hv0}) for $T_\star$ is simpler than solving the original equation. 

However, further simplification is possible by noting that
Eq.~(\ref{betaoverH}) implies a linear expansion of the Euclidean 
action around $t_\star$ so that the nucleation rate in Eq.~(\ref{Itb})
can be written as $\Gamma(t) = \Gamma_\star \, e^{\beta(t-t_\star)}$, where $\Gamma_\star \equiv \Gamma(t_\star)$. 
Since the integral is dominated by $t \simeq t_\star$~\cite{Turner:1992tz},
\be
I(t) \simeq 8\,\pi \,v_{\rm w}^3\,{\Gamma_\star \over \beta^4}\,e^{\beta(t-t_\star)} \,   .
\ee
Using this expression in $I(t_\star) = 1$, gives Eq.~(\ref{SETstar}).
Clearly, Eq.~(\ref{SETstar}) is much easier to solve if 
the Euclidean action is provided as a function of temperature.
We verified that solving Eq.~(\ref{SETstar})  gives results for $\a$ and $\beta/H_\star$ 
 in full agreement with those obtained by solving Eq.~(\ref{hv0}).

An even simpler and numerically efficient procedure employs the facts that $T_H$ and $T_\star$
are very close to each other with $T_H > T_\star$, and that the derivative of the Euclidean action is approximately constant in the interval
$[T_\star, T_H]$. Then, 
\be
S_E(T_\star) \simeq S_E(T_H) - S'_E(T_H) \, (T_H - T_\star) \,  ,
\ee
where $S_E(T_H)$ is given by  Eq.~(\ref{S3TH}). 
%We have verified
%that this simple, and fast, approximate procedure allows to calculate 
%$\a$ and $\b/H_\star$ very efficiently and with an accuracy within $10\%$. 

\subsection{Analytical expression for the Euclidean action in the general case}
\label{euclidapp}
%----------------------------------------------------------------------------

We provide expressions for the function $f(a_0,a_1)$ in Eq.~(\ref{fa0a1}) for the
Euclidean action obtained from the effective potential of Eq.~(\ref{gen}).

The two   vacua at $\langle \bar{\sigma}_1 \rangle = 0$ and $\langle \bar{\sigma}_1 \rangle = (\sqrt{4 a_1 +1}-1)/a_1$ become degenerate at the critical temperature
when the coefficients $a_0$ and $a_1$ satisfy the condition $a_0 = a_0^m(a_1)$, where
\begin{eqnarray} \label{eq:critical_condition}
a_0^m(a_1) \equiv a_1 \left[ \frac{2}{\sqrt{4a_1 +1}-1} -1 + 2\log\left(\frac{\sqrt{4a_1+1}+1}{4}\right) \right]\,. 
\end{eqnarray}
For a given $a_1$, the potential has a global minimum away from 0 only if $a_0 < a_0^m (a_1)$.
In particular for $a_1=0$, $a_0 < a_0^m (0) = 1$, and for $a_1=2$, $a_0 < a_0^m (0) = 0$. 
%This correlation is geometrically shown in Figs.~\ref{fig:a1_a0_highT} and \ref{fig:a1_a0_lowT}.
In the presence of the logarithmic term, a positive $a_1$ guarantees the stability of the vacuum, irrespective of the sign of $a_0$.
While $a_0< a_0^m(a_1)$ with $0<a_1<+\infty$ is an ideal condition to trigger the first-order phase transition, we require $a_0< a_0^m(a_1)-0.05$ to keep the numerical scan far away from the singularity at $a_0= a_0^m(a_1)$.

 Since a semi-analytical formula valid for the entire parameter space is challenging, we obtain two numerical fits
for the 3-dimensional Euclidean action in two separate regions:
\begin{eqnarray}
A &=& \{ (a_0, a_1) | 10^{-3} \leqslant  a_1 \leqslant 0.2 ,~
10^{-4} \leqslant a_0 < a_0^m (a_1) -0.05 \} \,, \nonumber\\
B &=& \{ (a_0, a_1) | 0.2\leqslant  a_1 \leqslant 10 ,~
-2 \leqslant a_0 < a_0^m (a_1) -0.05 \} \,.
\end{eqnarray}
For $(a_0,a_1) \in A$, we first fix the value of $a_1$. Then $a_0^m(a_1)$, the maximum value of $a_0$, is also fixed. We scan $a_0$ in the interval $(0, a_0^m(a_1)-0.05)$ and obtain a fit for $f(a_0,a_1)$. In region A, we fit a function of the form,
\begin{eqnarray}\label{eq:semi_analytical_1}
f(a_0,a_1) &=& f_0(a_1) + f_1(a_1) a_0 + \frac{f_2(a_1) a_0}{\alpha^m_0-a_0}
 + \frac{f_3(a_1) a_0}{(\alpha^m_0-a_0)^2} \,.
\end{eqnarray}
Varying $a_1$ in the interval $[10^{-3}, 0.2]$, we obtain the following numerical fits for the $f_i$'s:
\begin{eqnarray}
f_0(a_1) &=& \frac{5.31}{\left| a_1-2\right| ^{1.18}}+\frac{17.17}{(a_1-2)^2}+\frac{48.78}{(a_1-2)^3}
+\frac{70.59}{(a_1-2)^4} \,, \nonumber\\
f_1(a_1) &=& \widetilde{f}_0(a_1) e^{-1.91 a_1} \left(\frac{-0.52}{a_1-2}-\frac{1.47 a_1}{(a_1-2)^2}\right) \,, \nonumber\\
f_2(a_1) &=& \widetilde{f}_0(a_1) \frac{0.51 a_1^{3/2}+1.14 a_1-0.30 \sqrt{a_1}+0.64}{a_1^{3/2}+1} \,,\nonumber\\
f_3(a_1) &=& 0.04 a_1 \left| a_1-2\right| ^{0.012}+\frac{5.37 a_1}{(a_1-2)^2}+\frac{0.05 a_1}{(a_1-2)^3}  +\frac{0.02 a_1}{(a_1-2)^4} +0.30 \, .
\end{eqnarray}
In region B we express $f(a_0,a_1)$ in the form,
\begin{eqnarray} \label{eq:semi_analytical_2}
f(a_0,a_1) &=& \frac{g_1(a_1)}{(a_0^m(a_1)-a_0)^2} +
\frac{g_2(a_1)}{a_0^m(a_1)-a_0} 
+ \frac{g_3(a_1)}{(a_0^m(a_1)-a_0)^{1/2}}
+ \frac{g_4(a_1)}{(a_0^m(a_1)-a_0)^{1/4}} \nonumber\\
&&+ g_5(a_1) \log(a_0^m(a_1)-a_0) \,  ,
\end{eqnarray}
and obtain the following numerical fits for the $g_i$'s:
\begin{eqnarray}
g_1(a_1)&=&0.45 a_1+0.69\,,
\nonumber\\
g_2(a_1)&=&-0.40 \cos \left(41.16 -\frac{177.2}{\sqrt[4]{a'_1}}+\frac{229.2}{\sqrt{a'_1}}-\frac{120.6}{a'_1}\right)+2.76\,,
\nonumber\\
g_3(a_1)&=&-3.59 \cos \left(32.92 -\frac{161.6}{\sqrt[4]{a'_1}}+\frac{212.8}{\sqrt{a'_1}}-\frac{112.0}{a'_1}\right)+0.43\,,
\nonumber\\
\widetilde{g}_4(a_1)&=&3.46 \cos \left(26.40 -\frac{161.3}{\sqrt[4]{a'_1}}+\frac{211.9}{\sqrt{a'_1}}-\frac{110.4}{a'_1}\right) +0.71\,,
\nonumber\\
g_5(a_1)&=&-0.58 \cos \left(29.6 -\frac{146.5}{\sqrt[4]{a'_1}}+\frac{193.6}{\sqrt{a'_1}}-\frac{102.8}{a'_1}\right) -0.04\,,
\end{eqnarray}
where $a'_1 = a_1 +2$. 

\begin{figure}[t]
\begin{center}
  \psfig{file=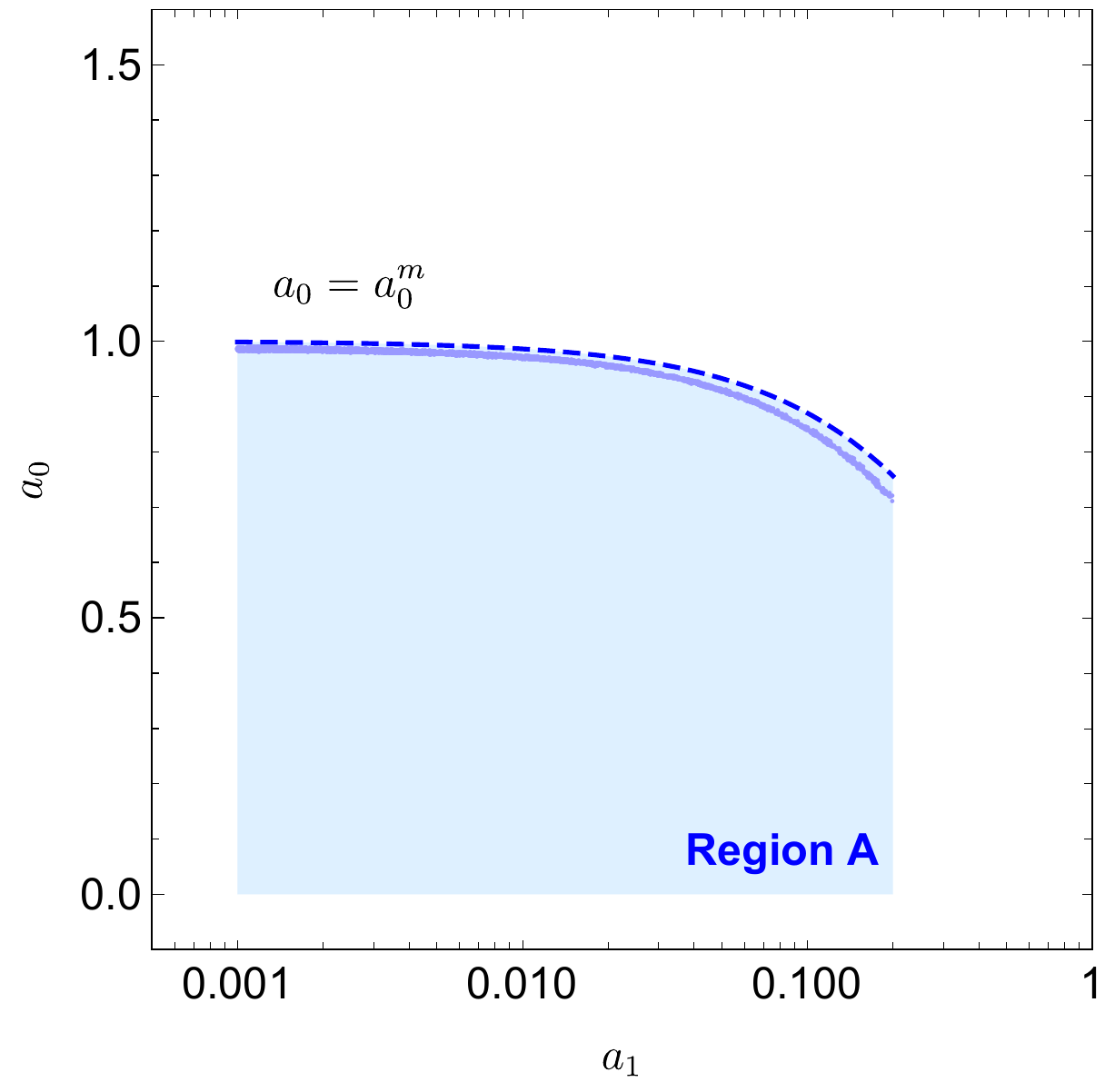,width=75mm}
        \psfig{file=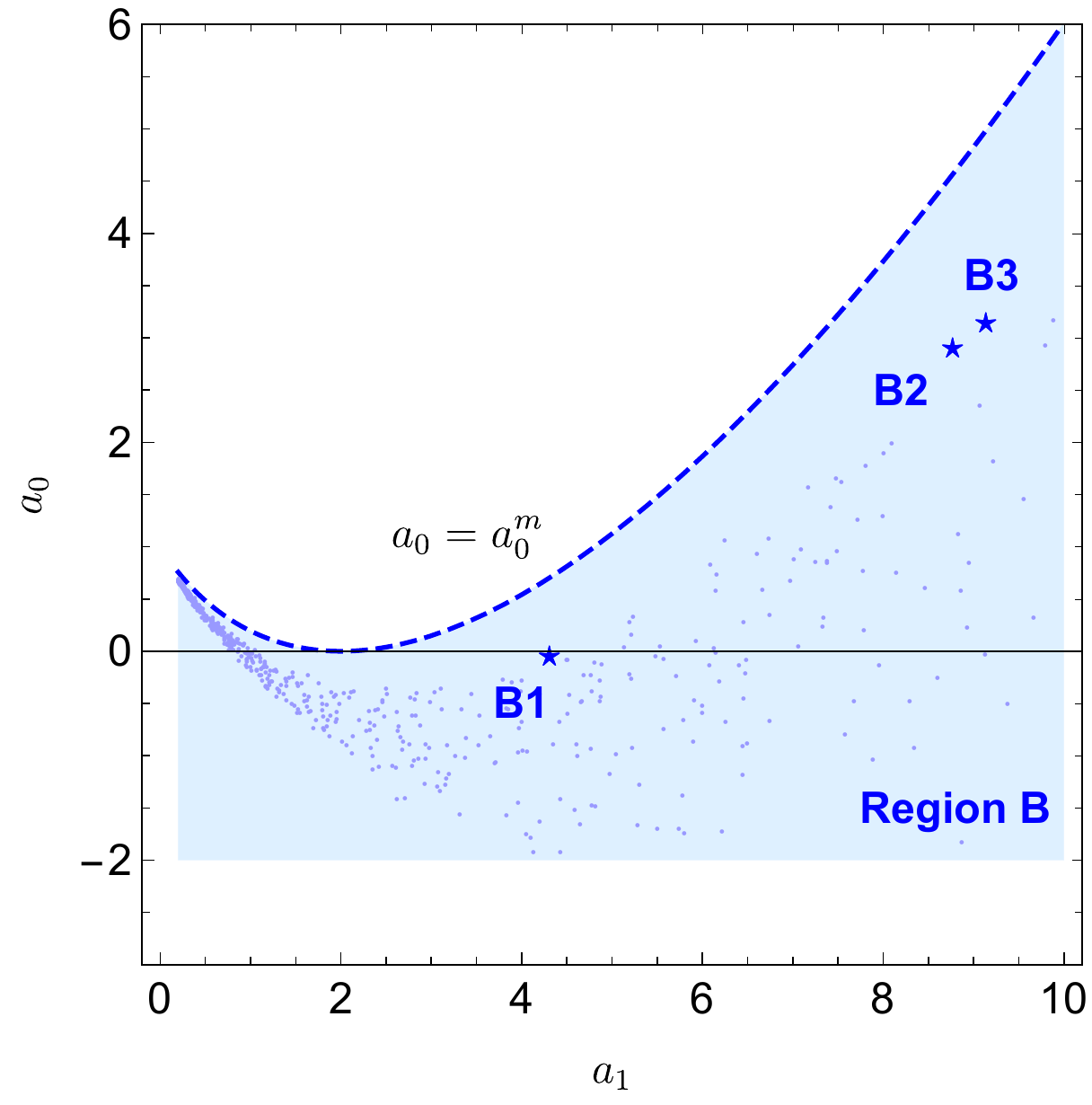,width=75mm}
        %\caption{resonant case}
        %\label{fig:gws}
\caption{Scatter plots in the $(a_0, a_1)$ plane in the high-scale scenario with logarithmic  field dependence
and no cubic term at zero temperature. The benchmark points in Table~\ref{tab:samples_highT} are marked with stars.}
\label{figa0a1}
\end{center}
\end{figure}
\begin{figure}[t!]
\begin{center}
        \psfig{file=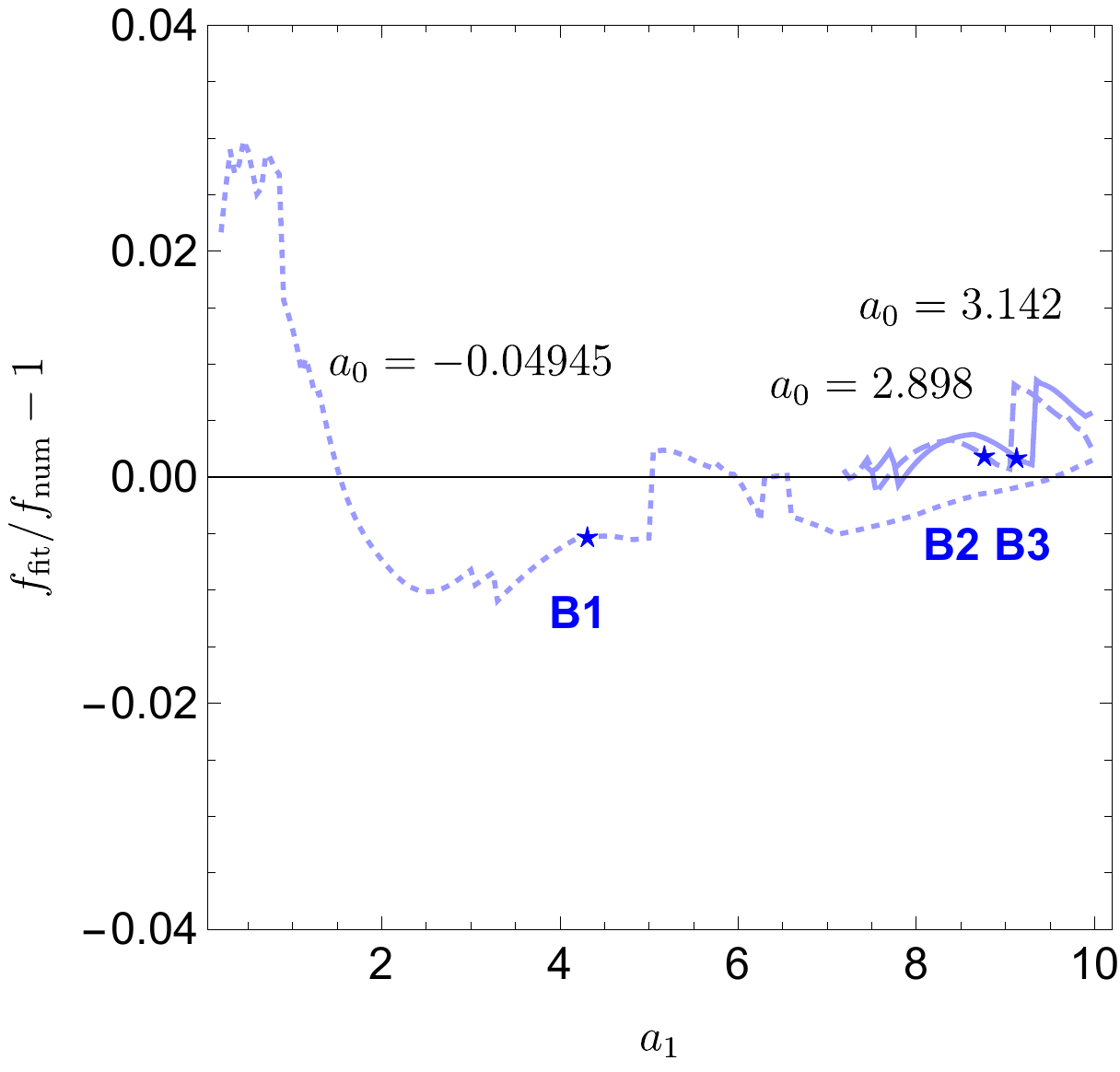,width=75mm}
        %\caption{resonant case}
        %\label{fig:gws}
\caption{Errors in the fit function for region B. Starred points correspond to values of $a_0$ 
for the benchmark points in Table~\ref{tab:samples_highT}.}
\label{fig:errors}
\end{center}
\end{figure}
%With these requirements, we have checked that our semi-analytical approximation matches with the numerical values in the scan with a precision about or %better than $5\%$. In particular, we show the relative errors of the semi-analytical approximation varying with $a_1$ with special fixed $a_0$ values. %All benchmark points (indicated as stars) have a precision around or better that $3\%$. 
Values of $a_0$ and $a_1$ in region $A$ (left panel) and $B$ (right panel) are shown in Fig.~\ref{figa0a1}. 
We verified that the analytical fits for $f(a_0,a_1)$ in regions A and B agree with  the numerical results within $5\%$. 
Figure~\ref{fig:errors} shows the errors for values of $a_0$ that correspond to benchmark points B1, B2 and B3.
We confirmed that in the limit $a_1 \ra 0$, we recover Eq.~(\ref{euclidean2}) with $a_0 = \widetilde{a}$.

%\newpage

\end{document}